\documentclass[12pt]{article}
\usepackage[T1]{fontenc}
\usepackage{amsmath,amssymb,amsthm,mathtools}
\usepackage{graphicx,enumitem,algorithm,algorithmic}
\usepackage{setspace}
\usepackage{comment}
\usepackage[round]{natbib}
\usepackage[hidelinks]{hyperref}

\allowdisplaybreaks
\newcommand{\E}{\mathop{{\mathrm{E}}}\nolimits}
\newcommand{\pr}{\mathop{{\mathrm{pr}}}\nolimits}
\newtheorem{theorem}{Theorem}
\newtheorem{lemma}{Lemma}
\newtheorem{corollary}{Corollary}
\newtheorem{proposition}{Proposition}
\theoremstyle{definition}
\newtheorem{definition}{Definition}

\theoremstyle{remark}
\newtheorem{remark}{Remark}

\begin{document}

	\title{{Active Hypothesis Testing under Computational Budgets}}
	
\author{Qi Kuang$^{1}$ \quad Bowen Gang$^{2}$ \quad Yin Xia$^{1}$\\[0.5em]
\small $^{1}$Department of Statistics and Data Science, Fudan University,\\
\small Shanghai 200433, China\\
\small $^{2}$Department of Mathematical and Statistical Sciences, University of Alberta,\\
\small Edmonton, Alberta T6G 2R3, Canada\\[0.3em]
\small \texttt{qkuang25@m.fudan.edu.cn}; \texttt{bgang@ualberta.ca}; \texttt{xiayin@fudan.edu.cn}}
	\date{}
	
	\maketitle

	\begin{abstract}
		
In large-scale hypothesis testing, computing exact $p$-values or $e$-values is often resource-intensive, creating a need for budget-aware inferential methods. We propose a general framework for active hypothesis testing that leverages inexpensive auxiliary statistics to allocate a global computational budget. For each hypothesis, our data-adaptive procedure probabilistically decides whether to compute the exact test statistic or a transformed proxy, guaranteeing a valid $p$-value or $e$-value under the stated assumptions while satisfying the exact budget constraint. We establish optimality for active $e$-values and develop optimality and admissibility theory for active $p$-values. Empirical results from simulations and two real-world applications, including a large-scale genome-wide association study and {a clinical prediction task using a large language model to impute a costly feature}, demonstrate that our framework improves statistical efficiency under fixed resource limits.
		
	\end{abstract}
	
	\noindent\textbf{Keywords:} Active Learning, Budget-aware Inference, Computational Constraints, $e$-value, Multiple Testing, $p$-value\par

	\section{Introduction}
The $p$-value and $e$-value \citep{vovk2021values, 10.1093/jrsssb/qkad085, ramdas2024hypothesis} are fundamental tools in statistical inference for quantifying evidence against a null hypothesis. While essential, their exact computation can be expensive due to costly experimental procedures or substantial computational demands. 
	This challenge creates a need for inferential methods that operate within a fixed budget. We propose a general framework for \textit{active hypothesis testing} that addresses this problem directly. Our approach leverages inexpensive and readily available auxiliary statistics, which are derived from cheaper data sources, prior knowledge, or predictive models, to manage a global computational budget. 
	For each hypothesis, a data-adaptive procedure probabilistically decides whether to compute the resource-intensive ``gold-standard'' statistic. When the exact statistic is not computed, a transformed version of the auxiliary statistic is used in its place, ensuring a valid test statistic for every hypothesis. This framework finds wide applicability across various domains, as illustrated by the following examples.

	\begin{enumerate}
		\renewcommand{\labelenumi}{(\alph{enumi})}
		
		\item {Powerful Prediction Model.} Consider a large, pre-trained model for predicting an outcome of interest \citep{Angelopoulos2023PPI,  JMLR:v24:23-0896, doi:10.1073/pnas.2322083121, kluger2025prediction, ji2025predictions}. Computing exact $p$-values or $e$-values requires actual outcomes, whose collection may be costly or delayed, whereas model predictions yield inexpensive proxy statistics. 
		In this setting, the proxy statistic acts as an auxiliary statistic to guide whether observing the true outcomes is worth the cost or delay.

		\item {Costly vs. Noisy Measurements.} In many scientific and industrial domains, a precise measurement is destructive, time-consuming, or financially expensive, while a cheaper but noisier measurement is often available from alternative sensors \citep{carroll1995measurement, fuller2009measurement, grace2021handbook, dunbar2022ensemble}. For instance, a full genetic assay is costly, but a simple biomarker measurement is not. Precise measurements support valid $p$-values or $e$-values, whereas noisy measurements provide potentially biased proxies.
		Our framework uses the noisy measurement to guide when to spend the budget on a precise measurement.
		
		\item {Multi-view Learning and Complementary Signals.} Different data views of the same phenomenon can provide complementary information. \citep{zhang2011multi, sun2013view,zhao201743}. A common setting involves two complementary data views. The first view, e.g., genomic or genotype data, is expensive to collect but supports valid computation of $p$-values or $e$-values. A second view, such as routine clinical measurements or gene expression data, is inexpensive to obtain but its null distribution is unknown, making it unsuitable for direct inferential procedures. Despite this limitation, the second view can provide substantial predictive signal.
		Our framework synthesizes these sources of information by using the cheap data view as a powerful proxy to strategically allocate the budget for computations on the resource-intensive view.
		
	\end{enumerate}

	Formally, we consider a setting with $N$ hypotheses, each associated with a resource-intensive test statistic that yields valid inference and a cheap auxiliary statistic that may be unreliable. Our objective is to construct a valid test statistic for \textit{every} hypothesis while ensuring that the number of costly computations {strictly adheres to} a predetermined global budget.
	
\subsection{Related Work}
	Prior work on incorporating auxiliary statistics in hypothesis testing has largely focused on improving statistical power, with limited consideration of computational budget constraints. These approaches typically leverage side information to prioritize hypotheses, and are often implemented through weighted multiple testing procedures, which can be interpreted either as re-weighting the $p$-values \citep{genovese2006false, ignatiadis2016data, liu2016new, barber2017p, xia2020gap, cai2022laws} or, equivalently, as adaptively adjusting the rejection thresholds \citep{lei2018adapt, zhang2019fast, li2019multiple, chao2021adapt, freestone2024semi}.
	Although these approaches improve power, they rely on the assumption that an exact $p$-value is available for every hypothesis and therefore do not tackle the fundamental challenge of high computational or experimental cost. Nevertheless, the weights or prioritization scores generated by some of these approaches can be leveraged as auxiliary statistics within our framework to inform efficient budget allocation.
    Similarly, two-stage multiple testing procedures use an inexpensive screening stage to filter out unpromising hypotheses \citep{zehetmayer2005two, aoshima2011two}. These methods, however, typically rely on a hard selection rule: hypotheses that fail the initial screening are discarded, and no formal inferential statements are made for them.
	
	A second line of research, which inspires our approach, is active learning, where information is queried selectively to improve efficiency \citep{cohn1996active, settles2009active, sener2018active, ren2021survey}.
	However, our goal is fundamentally different. The active learning literature, including recent work on acquiring gold-standard labels for statistical inference \citep{zhang2021M-estimators, zrnic2024active, cook2024semiparametric}, has primarily focused on optimizing the collection of {labelled} data for parameter estimation or model training. 
	While conceptually related, these methods aim to improve the efficiency of data collection, whereas our work focuses on the dynamic decision of whether to compute a test statistic itself.


The work most closely related to ours is the recently proposed proxy computing framework of \citet{xu2025active},  which employs probabilistic queries of exact test statistics to reduce expected computational costs. Their basic procedure, however, relies on a fixed test construction and makes query decisions independently for each hypothesis. As a result, it does not provide general optimality guarantees and the total computational cost remains stochastic. We generalize this framework by demonstrating that their construction is a special case of a broader class of valid active statistics. By characterizing this class, we establish the optimality and admissibility theory for active inference and replace the independent query mechanism with a budget-constrained global allocation.

\subsection{Our Contributions}
To address the aforementioned limitations, we develop a flexible and efficient framework for active hypothesis testing under a computational budget. {Our approach uses inexpensive auxiliary statistics to guide computational resource allocation, with the aim of improving statistical power while respecting budget constraints and maintaining statistical validity.}

Central to the procedure is a control function, guided by an auxiliary statistic, that probabilistically determines whether the expensive test statistic should be computed. When the exact statistic is not evaluated, a transformed version of the auxiliary statistic is used in its place, ensuring that a valid $p$-value or $e$-value is produced for every hypothesis. The framework requires only the availability of auxiliary information and a pre-specified budget, making it widely applicable across diverse scientific domains.

Our work makes several contributions. First, we establish a budgeted procedure that {guarantees the number of expensive computations exactly matches a user-specified limit on every run}. 
Second, our framework is model-free, imposing no distributional requirements on the auxiliary statistics. {This allows auxiliary statistics to be derived from predictive models, including models used to impute costly features.} Finally, we establish optimality for active $e$-values. Among active $p$-value constructions satisfying the proposed validity conditions, we obtain an optimal construction under independence and an admissible construction under general dependence. We further establish false discovery rate (FDR) control for the Benjamini--Hochberg (BH) procedure \citep{benjamini1995bh} applied to active $p$-values under conditional superuniformity and conditional PRDS of the original $p$-values given the auxiliary statistics, despite the dependence induced by exact-budget sampling. This positions our approach as a principled and theoretically sound method, rather than a heuristic.

\section{A Framework for Active Hypothesis Testing}\label{sec:budget}
\subsection{{Problem Formulation}}
	
	In many modern scientific applications, such as genomics, drug discovery, or large-scale A/B testing, the number of hypotheses to be tested far exceeds the available computational or experimental resources.\label{sec:formulation}
	Consider a set of $N$ null hypotheses, $\{H_{0,i}\}_{i=1}^N$. For each hypothesis $H_{0,i}$, we have access to two types of statistics:
	\begin{enumerate}
		\item A costly, valid test statistic, denoted generically by $X_i$. This represents the ``gold-standard" evidence and can be an $e$-value $E_i$ or a $p$-value $P_i$. The computation or acquisition of $X_i$ incurs a significant resource cost.
		\item An inexpensive auxiliary statistic, denoted by $X_i^a$. This statistic (e.g., $E_i^a$ or $P_i^a$) is readily available and is assumed to be informative about the exact statistic $X_i$, but it may not be statistically valid for formal inference on its own.
	\end{enumerate}
	
	Our primary objective is to generate a valid test statistic (an active $e$-value or active $p$-value) for \textit{every} hypothesis $i \in \{1, \dots, N\}$, while adhering to a pre-specified global budget. We assume that each computation of an expensive statistic $X_i$ incurs one unit of cost. The global budget, denoted by $n_b$ (where typically $n_b \ll N$), represents the total number of costly computations {allowed}. 
		Formally, let $C_i = \mathbb{I}(\text{statistic } X_i \text{ is computed})$ be an indicator variable for the decision to compute the expensive statistic for hypothesis $i$. The  budget constraint is then given by:
	\begin{equation}
		{\sum_{i=1}^N C_i \leq n_b.}
		\label{eq:budget_constraint}
	\end{equation}
	
To satisfy this global budget constraint while dynamically allocating resources to the most promising hypotheses, our framework employs hypothesis-specific control functions, $\{h_i\}_{i=1}^N$. The decision to compute $X_i$ is determined by the outcome of a Bernoulli trial with a success probability given by $h_i$, which may depend on the full vector of auxiliary statistics $\mathbf{X}^a = (X_1^a, \dots, X_N^a)$. 

The introduction of these control functions raises the central theoretical question of this work: how can one construct a test statistic that incorporates this probabilistic decision-making while rigorously preserving statistical validity? To answer this, we must first develop the fundamental building block at the level of a single hypothesis. 
We next define a new object, an ``active'' statistic, and establish its properties before demonstrating its use in the broader multiple testing setting. We develop this construction in two parallel frameworks, beginning with the active $e$-value.
	
	\subsection{Active $e$-value}\label{sec:active_e}
We begin by considering a single hypothesis.
Our goal is to construct an \textit{active $e$-value}, a statistic that leverages an inexpensive auxiliary statistic $E^a$ (nonnegative and without further distributional assumptions) to probabilistically decide whether to compute an exact, resource-intensive $e$-value $E$. Recall that a valid $e$-value is any non-negative random variable satisfying $\E _{H_0}[E] \le 1$, where larger values indicate stronger evidence against the null hypothesis.

The decision to compute $E$ is governed by a control function, $h: [0, \infty) \to [0, 1]$, which maps the observed value of $E^a$ to the probability of computing the exact $e$-value. This probabilistic rule leads to one of two outcomes for the final statistic, as formalized in the following definition.

	\begin{definition}[Active $e$-value]
		The active $e$-value is constructed as:
		\[
		E^{\mathrm{active}} = 
		\begin{cases} 
			a(E^a) & \text{if } U \ge h(E^a) \\
			b(E^a) \cdot E & \text{if } U < h(E^a),
		\end{cases}
		\]
		where $U \sim \text{Uniform}(0, 1)$ is independent of $(E^a, E)$, and $a(\cdot)$ and $b(\cdot)$ are non-negative functions to be designed such that $E^{\mathrm{active}}$ is a valid $e$-value.
		\label{def:active-e-value}
	\end{definition}
	
	\begin{remark}
		The choice of the multiplicative form $b(E^a)E$ is deliberate. It is designed to preserve the role of the exact $e$-value, $E$, which is typically a carefully constructed measure of evidence. This structure is intuitive and interpretable, as it simply re-scales the original evidence based on the auxiliary statistic $E^a$. We prefer this simple re-scaling to more complex transformations (e.g.,~$E^2$ or other nonlinear functions) that could obscure the relationship between the final statistic and the original $e$-value. Our framework also includes the active $e$-value proposed in \cite{xu2025active} as a special case, with a more detailed comparison provided in Section~\ref{sec:comparison_xu} of the supplement.

	\end{remark}
	
	The theoretical challenge we address next is to determine the conditions on $a(\cdot)$ and $b(\cdot)$ that ensure $E^{\mathrm{active}}$ preserves the $e$-value property, i.e., 
	$$\E [E^{\mathrm{active}}] = \E [a(E^{a}) \cdot (1- h(E^{a}))] + \E [b(E^{a})  h(E^{a})\cdot E]  \leq 1.$$

	A sufficient condition is to bound the first term by $\beta$ and the second by $1-\beta$, for some constant $\beta\in[0,1]$. The following theorem shows that such a decomposition is also necessary for validity over all admissible joint distributions of $(E^a,E)$.
	
	\begin{theorem}
		For $E^{\mathrm{active}}$ as defined in Definition~\ref{def:active-e-value}, given the control function $h(\cdot)$, the following two statements are equivalent: (1) $\E [E^{\mathrm{active}}] \le 1$ for all joint distributions of non-negative random variables $(E^a, E)$ with $\E [E] \le 1$; (2) There exists $\beta \in [0, 1]$ such that:
		$
		\sup_{x \ge 0} a(x)(1 - h(x)) \le \beta \text{ and } \sup_{x \ge 0} b(x)h(x) \le 1-\beta.
		$
	
		\label{thm:e-value-characterization}
	\end{theorem}
	
	The characterization in Theorem~\ref{thm:e-value-characterization} directly informs the optimal design of the functions $a(\cdot)$ and $b(\cdot)$, 
	as demonstrated in the following corollary.
	
	\begin{corollary}
		For any given $\beta \in [0, 1]$ and control function $h(\cdot)$, set:
		\[
		a(x) = \frac{\beta}{1 - h(x)} \quad \text{and} \quad b(x) = \frac{1 - \beta}{h(x)}.
		\]
		Then $E^{\mathrm{active}}$ is a valid $e$-value and achieves the tight bound in Theorem~\ref{thm:e-value-characterization}. In other words, for a fixed $h(\cdot)$ and $\beta$, this construction is optimal in the sense that it is point-wise the largest possible, thus maximizing the resulting $e$-value while preserving validity.
		\label{cor:optimal-construction}
	\end{corollary}
	
	Thus, for a given $\beta$, the active $e$-value construction that is optimal for a fixed control function takes the following form:
	\begin{equation}\label{active-e}
		E^{\mathrm{active}} = 
		\begin{cases} 
			\dfrac{\beta}{1 - h(E^a)} & \text{if } U \ge h(E^a) \\
			\dfrac{1 - \beta}{h(E^a)} \cdot E & \text{if } U < h(E^a).
		\end{cases}
	\end{equation}

	\subsection{Active $p$-value}\label{sec:active_p}
	We now develop an analogous framework for active $p$-values, extending the core principles established for $e$-values. The conceptual setup mirrors the $e$-value case: for a given hypothesis, we have access to an exact, valid $p$-value $P$, and an inexpensive auxiliary statistic $P^a$ taking values in $[0, 1]$ without further distributional assumptions. We recall that a valid $p$-value $P$ is a random variable satisfying the super-uniformity property under the null hypothesis $H_0$: $\pr _{H_0}(P \le s) \le s$ for all $s \in [0, 1]$.
	
	Mirroring our approach for $e$-values, the decision to compute the expensive $p$-value is governed by a control function, $h(\cdot): [0, 1] \to [0, 1]$. This function maps the observed value of the auxiliary statistic $P^a$ to the probability of computing the expensive $p$-value, leading to the following definition.
	\begin{definition}[Active $p$-value]\label{def:active-p-value}
		The active $p$-value, $P^{\mathrm{active}}$, is constructed as follows:
		\[
		P^{\mathrm{active}}= 
		\begin{cases} 
			a(P^a) & \text{if } U \ge h(P^a) \\
			b(P^a) \cdot P & \text{if } U < h(P^a),
		\end{cases}
		\]
		where $U \sim \mathrm{Uniform}(0, 1)$ is independent of $(P^a, P)$, and the non-negative measurable functions $a(\cdot)$ and $b(\cdot)$ must be chosen to ensure that $P^{\mathrm{active}}$ is a valid $p$-value.
	\end{definition}
	
	\begin{remark}\label{rem:p-value-form}
		Similarly to the $e$-value case, the multiplicative form $b(P^a)P$ is a deliberate choice. It preserves the original structure of the exact $p$-value $P$, which is often carefully constructed for high power, by simply re-scaling it \citep[e.g.,][]{barber2017p, li2019multiple, xia2020gap, cai2022laws}. Moreover, our framework encompasses the active $p$-value proposed in \cite{xu2025active} as a special case. Further discussions are provided in Section~\ref{sec:comparison_xu} of the supplement.
	\end{remark}
	
	The theoretical challenge is to determine the conditions on $a(\cdot)$ and $b(\cdot)$ that ensure $P^{\mathrm{active}}$ is a valid $p$-value. This requires that for all $s \in [0, 1]$,
	\begin{align}\label{eq:p-validity}
		\pr (P^{\mathrm{active}}\le s) = \E \left[ (1-h(P^a))\mathbb{I}\{a(P^a) \le s\} + h(P^a)\mathbb{I}\{b(P^a)P \le s\} \right] \le s.
	\end{align}

	To satisfy this validity condition, we again take a decomposition approach. For a given $\beta \in {(0, 1)}$, we can ensure the total probability is bounded by $s$ if we require that the two terms in the sum are bounded by $\beta s$ and $(1-\beta)s$ respectively:
	\begin{align}
		\E \left[(1 - h(P^a))\mathbb{I}\{a(P^a) \le s\}\right] &\le \beta s \label{eq:p-cond-1}, \\
		\E \left[h(P^a)\mathbb{I}\{b(P^a)P \le s\}\right] &\le (1-\beta)s, \label{eq:p-cond-2}
	\end{align}
	for all $s \in [0, 1]$. To make the resulting test statistic  powerful, we must choose $a(\cdot)$ and $b(\cdot)$ to make the active $p$-value as small as possible. This requires minimizing both of its potential outcomes, $a(P^a)$ and $b(P^a)P$, subject to their respective validity constraints.
	
	\begin{remark}\label{rem:decomposition}
The separate constraints \eqref{eq:p-cond-1} and \eqref{eq:p-cond-2} are sufficient, but not necessary, to ensure the active $p$-value satisfies the super-uniformity condition in \eqref{eq:p-validity}. A counterexample is provided in Section~\ref{sec:counterexample-decomposition} of the supplement. This contrasts with the $e$-value case in Theorem~\ref{thm:e-value-characterization}. While our approach imposes a stronger condition than strictly required, it furnishes a tractable framework for constructing a broad class of valid active $p$-values.
	\end{remark}
	We next turn to the optimal form of $a(\cdot)$ from Condition \eqref{eq:p-cond-1}. Since $a(P^a)$ serves as a component of the $p$-value, only values where $a(x) \le 1$ are meaningful for inference. {To obtain smaller active $p$-values,} we seek the point-wise smallest function $a(\cdot)$. The following theorem identifies this optimal choice.

\begin{theorem}\label{prop:a}
		Given $\beta$ and $h(\cdot)$, if $a(\cdot)$ satisfies \eqref{eq:p-cond-1} for all distributions of $P^a\in [0,1]$ and all $s\in [0,1]$, then $a(x)\geq  (1-h(x))/\beta$ whenever $a(x)\leq 1$. 
		Consequently, wherever \((1-h(x))/\beta\le1\), the choice \(a(x)=(1-h(x))/\beta\) is pointwise smallest among functions satisfying \eqref{eq:p-cond-1}.
	\end{theorem}
	
In contrast, the optimal choice of $b(\cdot)$ under Condition~\eqref{eq:p-cond-2} is more nuanced, as it is governed by the joint distribution of $P$ and $P^a$. For instance, the choice $b(q) = h(q)/(1-\beta)$, which is analogous to the optimal $e$-value construction, fails to satisfy Condition~\eqref{eq:p-cond-2} under general dependence (a counterexample is provided in Section~\ref{sec:counterexample} of the supplement). This distinction motivates the need for separate constructions depending on the dependency structure, which we formalize in the following theorem.
	\begin{theorem}\label{thm:active_p_admissible}
	For fixed $h(\cdot)$ and $\beta$, we have
	\begin{enumerate}
		\item If $P$ and $P^a$ are independent, the point-wise smallest $b(\cdot)$ that satisfies \eqref{eq:p-cond-2} is:
		$
		b(x) = {h(x)}/{(1 - \beta)}.
		$
		\item Under general dependence, an admissible choice for $b(\cdot)$ that satisfies \eqref{eq:p-cond-2} is:
		$
		b(x) = {\sup_y h(y)}/{(1 - \beta)} \cdot \mathbb{I}(h(x) > 0).
		$
		Here, admissibility means that no other valid function $\tilde{b}(\cdot)$ can strictly dominate this choice, i.e., there is no $\tilde{b}(\cdot)$ satisfying \eqref{eq:p-cond-2} such that $\tilde{b}(x) \le b(x)$ for all $x$ and $\tilde{b}(x_0) < b(x_0)$ for at least one point $x_0$.
	\end{enumerate}
\end{theorem}

	Theorems \ref{prop:a} - \ref{thm:active_p_admissible} lead to the construction of the active $p$-value. In the following text, the term ``active $p$-value'' refers to one of these two forms, depending on the dependence between $P$ and $P^a$.
	\par\smallskip\noindent\textit{Under Independence}\enspace 
	When the exact $p$-value $P$ and the auxiliary statistic $P^a$ are independent, the active $p$-value takes the following form:
	\begin{equation}	\label{eq:p-value-indep}
		P^{\mathrm{active}}= 
		\begin{cases} 
			\dfrac{1 - h(P^a)}{\beta} & \text{if } U \ge h(P^a) \\[1em]
			\dfrac{h(P^a)}{1 - \beta} \cdot P & \text{if } U < h(P^a).
		\end{cases}
	\end{equation}
	
	\par\smallskip\noindent\textit{Under General Dependence}\enspace 
	To guarantee validity under arbitrary dependence between \(P\) and \(P^a\), we use \(\sup_x h(x)\) in place of \(h(P^a)\). The resulting active $p$-value is:
	\begin{equation}	\label{eq:p-value-general}
		P^{\mathrm{active}}= 
		\begin{cases} 
			\dfrac{1 - h(P^a)}{\beta} & \text{if } U \ge h(P^a) \\[1em]
			\dfrac{\sup_x h(x)}{1 - \beta} \cdot P & \text{if } U < h(P^a).
		\end{cases}
	\end{equation}
	
	A direct comparison of the two forms in \eqref{eq:p-value-indep} and \eqref{eq:p-value-general} reveals the trade-off between statistical efficiency and robustness. When independence can be assumed, the resulting active $p$-value is smaller (and thus more powerful), as $h(P^a) \le \sup_x h(x)$. The construction for general dependence pays a price in statistical power to guarantee validity in a wider range of scenarios.

	\subsection{Admissibility and the Choice of Control Parameters}\label{sec:hbeta}
	The active statistic constructions in \eqref{active-e}, \eqref{eq:p-value-indep} and \eqref{eq:p-value-general} depend on the choice of the control function $h(\cdot)$ and the hyperparameter $\beta$. 
	While recent literature \citep{xu2025active} has introduced specific functional forms for active statistics, a rigorous theoretical evaluation of whether these or any other choices are optimal has remained absent. This  naturally raises a question: does a universally optimal configuration actually exist? That is, can we identify a specific function $h$ and parameter $\beta$ that yield a strictly more powerful test against \emph{all} alternatives?
		To answer this question and address the gap in prior work, we provide an in-depth theoretical investigation into the \emph{admissibility} of active statistics.  We begin by formally defining statistical domination and admissibility within our framework. Intuitively, one active statistic dominates another if it is always ``better'', which means yielding a larger $e$-value or smaller $p$-value.

	
	
	\begin{definition}[Domination and Admissibility]
		Let $X^{\text{active}}_{h,\beta}$ denote an active statistic (either an active $e$-value or $p$-value) constructed using control function $h$ and parameter $\beta$.
		All comparisons use the same input statistics and the same sampling variable $U$, independent of those statistics.
		We say that $X^{\text{active}}_{h,\beta}$ {dominates} $X^{\text{active}}_{h',\beta'}$ if its statistic is uniformly no worse, with strict improvement for at least one valid input law. Formally:
		\begin{enumerate}
	  \item {For $p$-values:} For any valid $p$-value $P$ and auxiliary statistic $P^a \in [0,1]$, the inequality 
	$
	\min\{1, P^{\mathrm{active}}_{h,\beta}\} \le \min\{1, P^{\mathrm{active}}_{h',\beta'}\}
	$
	holds almost surely, and strict inequality holds with positive probability for at least one valid input pair.
	
	\item {For $e$-values:} For any valid $e$-value $E$ and auxiliary statistic $E^a\geq 0$, the inequality 
	$
	E^{\text{active}}_{h,\beta} \ge E^{\text{active}}_{h',\beta'}
	$
	holds almost surely, and strict inequality holds with positive probability for at least one valid input pair.
		\end{enumerate}
		An active statistic is {admissible} if it is not dominated by any other active statistic. We say that a choice of $h(\cdot)$ or $\beta$ is admissible if the resulting active statistic is admissible.
	\end{definition}

The following propositions establish a key theoretical property of our framework. No single choice of control parameters is universally superior.

\begin{proposition}
\label{prop:no-optimal-h}
	Fix $\beta \in (0, 1)$. No single control function $h(\cdot)$ uniformly dominates all others. 
	\begin{enumerate}
		\item For active $e$-values, every choice of $h(\cdot)$ is admissible.
		\item For active $p$-values under independence, as in \eqref{eq:p-value-indep}, $h$ is admissible if and only if $h(\cdot)>0$. Under general dependence, as in \eqref{eq:p-value-general}, $h$ is admissible if and only if either $h(\cdot)>1-\beta$, or $h(\cdot)\equiv t$ for some $t\in(0,1]$.
	\end{enumerate}
\end{proposition}

\begin{proposition}
\label{prop:no-optimal-beta}
	Fix a non-trivial $h$ that is not identically $0$ or $1$: 
	\begin{enumerate}
		\item For active $e$-values, every $\beta\in(0,1)$ is admissible.
		\item For active $p$-values, $\beta\in(0,1)$ is admissible if and only if $1-\beta<h(x)<1$ for some $x\in[0,1]$. Thus, if $\sup\{h(x):h(x)<1\}=1$, every $\beta\in(0,1)$ is admissible.
	\end{enumerate}
\end{proposition}
The choice of \(\beta\) involves a trade-off between the two branches. A larger \(\beta\) yields stronger evidence when the exact statistic is not queried, whereas a smaller \(\beta\) yields stronger evidence when it is queried. We use \(\beta=0.5\) as a default to balance the two branches.

Finally, while the results in this section focus on a single hypothesis for clarity, they extend naturally to the multivariate setting where $h_i$ depends on the full vector of auxiliary statistics $\mathbf{X}^a$. We provide the formal extension and proofs of multivariate admissibility in Section~\ref{sec:admiss-multi} of the supplement.

	\section{Hypothesis Testing under Budget Constraint}\label{sec:budget2}
\subsection{A Normalized Allocation Scheme}
The admissibility results in Section~\ref{sec:hbeta} show that no single choice of control functions is uniformly best. We therefore use auxiliary statistics to guide budget allocation, with the aim of achieving high statistical power under the given budget. We now return to the problem formulated in Section \ref{sec:formulation} and present such a strategy.

\label{sec:allocation_scheme}

To connect the global budget $n_b$ to the individual decision probabilities $\{h_i\}$, we introduce the concept of a \textit{utility function}, $u_i(\cdot)$. For each hypothesis $i$, the utility function $u_i: \mathcal{X}^a \to \mathbb{R}_{\ge 0}$ maps the auxiliary statistic $X_i^a$ to a non-negative score that quantifies the ``desirability" of computing the exact statistic $X_i$. A larger value of $u_i(X_i^a)$ indicates a higher priority for allocation of the computational budget.

Given a set of utility functions $\{u_i\}_{i=1}^N$, we define the control function for each hypothesis via a normalized allocation scheme:
\begin{equation}
	h_i(\mathbf{X}^a) = n_b \cdot \frac{u_i(X_i^a)}{\sum_{j=1}^N u_j(X_j^a)}.
	\label{eq:h_i_definition}
\end{equation}
By construction, this ensures the exact sum constraint $\sum_{i=1}^N h_i(\mathbf{X}^a) = n_b$.
\subsection{Guidance on Selecting the Utility Functions}
\label{sec:utility_function_guidance}

Principled strategies for selecting the functional form of $u_i$ can lead to substantial gains. The core idea is to encode prior knowledge about the relationship between the auxiliary and exact statistics into the functional form of $u_i$.

 In most applications, the auxiliary statistic $X_i^a$ exhibits a directional relationship with the strength of the evidence against the null. We classify this into two cases:
\begin{enumerate}
	\item {Direct Signal:} 
	A signal is considered \textit{direct} when larger values of $X_i^a$ are more indicative of the alternative hypothesis. For example, a large $E_i^a$ may serve as a proxy for a large exact $e$-value $E_i$. For direct signals, a non-decreasing utility function $u_i(\cdot)$ should be chosen. A natural default choice is the identity function, $u_i(x) = x$.
	\item {Inverse Signal:} 
	 A signal is \textit{inverse} when smaller values of $X_i^a$ are more indicative of the alternative hypothesis (e.g., a small $P_i^a$ serving as a proxy for an exact $p$-value $P_i$). For inverse signals, a non-increasing utility function is appropriate. A standard choice is $u_i(x) = 1/(x + \epsilon)$, where $\epsilon > 0$ is a small constant for numerical stability.
\end{enumerate}

When the normalized allocation in \eqref{eq:h_i_definition} exceeds one, we recursively compress the utilities using $u_i\gets\log(1+u_i^{1/k})$, with $k=1,2,\ldots$, and renormalize. Algorithm~\ref{alg:unified_framework} handles boundary cases separately and guarantees termination, yielding query probabilities $h_i\in[0,1]$ that sum to $n_b$.

This utility selection strategy creates a strong synergy. Consider the active $e$-value construction \eqref{active-e}. Under the alternative, a promising auxiliary statistic (e.g., a large $E_i^a$ in the direct signal case) will produce a large utility $u_i(E_i^a)$, which in turn increases its control value $h_i(E_i^a)$. This yields two benefits:
\begin{enumerate}
	\item It increases the probability of computing the gold-standard $e$-value $E_i$, which is also expected to be large.
	\item In the event that $E_i$ is not computed, the resulting auxiliary-based statistic, $\beta / (1 - h_i(E_i^a))$, is also larger, thereby amplifying the evidence from the auxiliary statistic itself.
\end{enumerate}
This allocation strategy prioritizes hypotheses with promising auxiliary statistics.

\subsection{Budgeted Active Inference Algorithm}
\label{sec:complete_algorithm}
Next, a central technical challenge is to ensure that the total number of expensive computations \emph{exactly} equals $n_b$ on every run. Unlike previous methods \citep{xu2025active} that rely on independent coin flips which result in random, unpredictable budget utilization, our framework requires a dependent sampling mechanism that correlates the decisions across all $N$ hypotheses to ensure strict budget adherence. Formally, we seek to sample binary indicators $C_1, \dots, C_N \in \{0, 1\}$ conditionally on $\mathbf{X}^a$ such that they satisfy two conditions simultaneously: valid marginal selection probabilities, meaning $C_i \mid \mathbf{X}^a \sim \mathrm{Bernoulli}(h_i(\mathbf{X}^a))$ for each $i$; and exact global budget adherence, meaning $\sum_{i=1}^N C_i = n_b$. While the theoretical existence of such a joint distribution is guaranteed by \citet{chen2022joint}, this existence result does not directly yield a practical sampling algorithm. To address this, the next proposition provides an explicit construction of $C_1,\ldots, C_N$ that satisfies these conditions.
\begin{proposition}
	\label{prop:systematic_sampling}
Suppose $p_1, \dots, p_N \in [0, 1]$ satisfy $\sum_{i=1}^N p_i = n_b \in \mathbb{N}$. Let $S_i = \sum_{j=1}^i p_j$ for $i = 1, \dots, N$, with $S_0 = 0$ and $U \sim \mathrm{Uniform}(0, 1)$. Define
	\begin{equation}
		C_i = \lfloor S_i - U \rfloor - \lfloor S_{i-1} - U \rfloor, \quad i = 1, \dots, N.
		\label{eq:systematic_sampling}
	\end{equation}
	Then marginally $C_i\sim \text{Bernoulli}(p_i)$ for all $i$, and $\sum_{i=1}^N C_i = n_b$.
\end{proposition}

 We are now ready to present the complete algorithm for budgeted active inference. 
The procedure, detailed in Algorithm~\ref{alg:unified_framework}, takes as input the auxiliary statistics, a global exact budget, the hyperparameter $\beta$, and user-specified utility functions. It returns a valid test statistic for every hypothesis and rigorously adheres to the constraints. 

\begin{algorithm}[!htbp]
	\caption{A Unified Algorithm for Budgeted Active Inference}
	\label{alg:unified_framework}
	\begin{algorithmic}[1]
		\renewcommand{\algorithmicrequire}{{Input:}}
		\REQUIRE 
		Auxiliary statistics $\{X_i^a\}_{i=1}^N$, 
		exact global budget $n_b\in\{0,\ldots,N\}$, 
		hyperparameter $\beta \in (0,1)$, 
		user-specified base utility functions $\{u_i(\cdot)\}_{i=1}^N$.
		\STATE {Step 1: Normalized Allocation}
		{
        \STATE Let $u_i=u_i(X_i^a)$ for $i=1,\ldots,N$, and let $m=|\{i:u_i>0\}|$.
        \IF{$n_b=0$ or $n_b=N$}
            \STATE Set $h_i=n_b/N$ for all $i$.
        \ELSIF{$m\le n_b$}
            \STATE Set $h_i=1$ if $u_i>0$ and $h_i=(n_b-m)/(N-m)$ otherwise.
        \ELSE
            \STATE Set $k=1$ and $h_i=n_bu_i/\sum_j u_j$ for all $i$.
            \WHILE{$\max_i h_i>1$}
                \STATE Update $u_i\gets\log(1+u_i^{1/k})$ for all $i$.
                \STATE Recompute $h_i=n_bu_i/\sum_j u_j$ for all $i$, and set $k\gets k+1$.
            \ENDWHILE
        \ENDIF
        }
        
        \STATE {Step 2: Exact-Sum Dependent Sampling}
		\STATE Compute cumulative limits $S_i = \sum_{j=1}^i h_j$ (with $S_0 = 0$).
		\STATE Sample $U \sim \mathrm{Uniform}(0, 1)$, independently of the input statistics.
		
		\FOR{each hypothesis $i = 1, \dots, N$}
		\STATE Set indicator boolean $C_i = \lfloor S_i - U \rfloor - \lfloor S_{i-1} - U \rfloor$.
		\IF{$C_i = 1$}
		\STATE Compute the exact primary statistic $X_i$.
		\STATE Construct the active statistic based on $X_i$:
		\begin{enumerate}[noitemsep, topsep=0pt, parsep=0pt, label={(\roman*)}]
			\item {For $e$-values:} $X_i^{\mathrm{active}} \gets {(1-\beta)/(h_i)}\,X_i$.
			\item {For $p$-values:} Set $X_i^{\mathrm{active}} \gets \min(1, b_i\,X_i)$, where the scaling factor $b_i$ is:
			      \begin{enumerate}[noitemsep, topsep=0pt, parsep=0pt, label={(\roman*)}]
			      	\item $b_i = {(h_i)/(1-\beta)}$ (under independence, i.e. $P_i \perp \mathbf{P}^a$),
					
			      	\item $b_i = {(\min(1, \sup_{\mathbf{y} \in \mathbb{R}^N} h_i(\mathbf{y})))/(1-\beta)}$ (under general dependence).
			      \end{enumerate}
		\end{enumerate}
		\ELSE
		\STATE Construct the active statistic without $X_i$:
		\begin{enumerate}[noitemsep, topsep=0pt, parsep=0pt, label={(\roman*)}]
			\item {For $e$-values:} $X_i^{\mathrm{active}} \gets {(\beta)/(1 - h_i)}$.
			\item {For $p$-values:} $X_i^{\mathrm{active}} \gets \min(1, {(1 - h_i)/(\beta)})$.
		\end{enumerate}
		\ENDIF
		\ENDFOR
		\STATE {Return:} The set of active statistics $\{X_i^{\mathrm{active}}\}_{i=1}^N$.
	\end{algorithmic}
\end{algorithm}

{

\subsection{False Discovery Rate Control}
\label{sec:budget_fdr}
The active statistics returned by Algorithm~\ref{alg:unified_framework} can be used for multiple testing via BH or e-BH. While e-BH controls FDR under arbitrary dependence \citep{wang2022ebh}, applying BH to the active $p$-values requires additional justification. In particular, all active $p$-values depend on the auxiliary statistics $\mathbf P^a$. Their dependence structure therefore reflects not only the dependence among the original $p$-values, but also the dependence involving $\mathbf P^a$, and can be considerably more complex. Thus, PRDS of the original $p$-values does not by itself imply PRDS of the active $p$-values or guarantee FDR control for BH applied to them. The following theorem establishes FDR control for BH applied to the active $p$-values under the stated conditions.

\begin{theorem}
\label{thm:active_bh_conditional}
Let $\mathcal H_0$ be the set of true null hypotheses, and fix $\alpha,\beta\in(0,1)$. Consider Algorithm~\ref{alg:unified_framework}. Suppose that, conditional on $\mathbf P^a$, almost surely:
\begin{enumerate}
\item Each original null $p$-value is valid: $\pr (P_i\le t\mid\mathbf P^a)\le t$ for all $i\in\mathcal H_0$ and $t\in[0,1]$.
\item The full vector $\mathbf P$ is PRDS on $\mathcal H_0$: for every coordinatewise increasing measurable set $D\subseteq[0,1]^N$ and every $i\in\mathcal H_0$, the function $t\mapsto\pr (\mathbf P\in D\mid P_i\le t,\mathbf P^a)$ is nondecreasing wherever the conditioning event has positive conditional probability.
\end{enumerate}
Using either queried scaling $b_i=h_i/(1-\beta)$ or $b_i=\min\{1,\sup_{\mathbf y}h_i(\mathbf y)\}/(1-\beta)$, BH applied at level $\alpha$ to the active $p$-values returned by Algorithm~\ref{alg:unified_framework} controls the FDR at level $\alpha$.
\end{theorem}

 Conditional superuniformity and conditional PRDS therefore ensure FDR control for the active $p$-values. Independence of the original and auxiliary vectors gives a sufficient condition for these conditional properties, as stated below.

\begin{corollary}
\label{cor:active_bh_independent}
Suppose that $\mathbf P\perp\mathbf P^a$, each $P_i$ is a valid $p$-value for $i\in\mathcal H_0$, and the full vector $\mathbf P$ is PRDS on $\mathcal H_0$. BH applied to the active $p$-values controls FDR at level $\alpha$.
\end{corollary}

}

\section{Numerical Experiments}\label{sec:simulations}
\subsection{{Competing Methods and Evaluation Metrics}}
We conduct numerical simulations to evaluate our budgeted active inference framework, comparing its statistical power and efficiency against several baselines under a fixed budget.
Across all experiments we use the function $u_i(x)=x$ for direct signals and $u_i(x)=1/(x+\epsilon)$ for inverse signals as our base utility functions. Any necessary range compression to bound extreme values is automatically handled by the adaptive constraint loop built within Algorithm~\ref{alg:unified_framework}.

We compare Algorithm ~\ref{alg:unified_framework}, referred to as ``\texttt{Active-Default}", with the following methods:

\begin{enumerate}[noitemsep, topsep=0pt, parsep=0pt]
\item {\texttt{ALL} (Oracle).} This non-budgeted oracle method computes the exact statistic ($E_i$ or $P_i$) for all $N$ hypotheses. It serves as a full-data power comparator at a fixed cost of $N$ queries.

\item {\texttt{Random}.} 
A simple baseline that adheres to the budget by selecting a uniform random subset of $n_b$ hypotheses to query. For any hypothesis that is not selected, its active statistic is set to the non-informative value of 1.

\item {\texttt{Xu} \citep{xu2025active}, basic construction.} 
This method makes an independent probabilistic decision for each hypothesis. A Bernoulli trial $T_i \sim \mathrm{Bernoulli}(p_i)$ determines whether to compute the expensive statistic, where the probability $p_i$ is a function of the auxiliary statistic and a hyperparameter $\beta$.
For $e$-values, the query probability is $p_i = \max\{0, 1-\beta/E_i^a\}$, and the final statistic is $E_i^{\mathrm{active}} = (1-T_i)E_i^a + T_i(1-\beta)E_i$.
For $p$-values, $p_i = \max\{0, 1-\beta P_i^a\}$, and the final statistic is $P_i^{\mathrm{active}} = (1-T_i)P_i^a + T_i(1-\beta)^{-1}P_i$.
Crucially, because these decisions are made independently for each hypothesis, the total number of queries $\sum_i T_i$ is a random variable and is not constrained by a pre-specified global budget.

\item {\texttt{Active-Xu}.} It uses the utility function implied by \texttt{Xu} (e.g. $u_i(x) = \max(1 - \beta/x, 0)$ for e-values), but embeds it within our global budget allocation framework.

\end{enumerate}
To evaluate the output statistics, we apply the e-BH procedure to $e$-values and the BY procedure to $p$-values at an FDR level of $\alpha = 0.1$, as both are robust to arbitrary dependence structures. Unless otherwise specified, all active methods use the hyperparameter $\beta=0.5$.
Our evaluation {centres} on the trade-off between statistical power and computational cost. We adopt the following standard notation: $\mathcal{H}_0$ and $\mathcal{H}_1$ denote the sets of true null and non-null hypotheses, respectively, with $|\mathcal{H}_1| = N_1 > 0$. For a given method, $\mathcal{R}$ is the set of rejected hypotheses.

\par\smallskip\noindent{\textit{Statistical Validity and Power.}}\enspace 
\begin{enumerate}[noitemsep, topsep=0pt, parsep=0pt, label={(\roman*)}]
	\item {FDR:} The expected proportion of false discoveries, defined as ${\text{\textsc{fdr}}} = \E [V / \max(|\mathcal{R}|, 1)]$, where $V = |\mathcal{R} \cap \mathcal{H}_0|$. All methods are expected to satisfy ${\text{\textsc{fdr}}} \le \alpha$.
	\item {True Positive Rate (TPR):} The expected proportion of true non-nulls correctly rejected, defined as $\text{Power} = \E [S / N_1]$, where $S = |\mathcal{R} \cap \mathcal{H}_1|$.
\end{enumerate}

\par\smallskip\noindent{\textit{Budget-Aware Performance.}}\enspace 
Since all methods control FDR, our primary comparison hinges on the efficient use of the computational budget.
\begin{enumerate}[noitemsep, topsep=0pt, parsep=0pt, label={(\roman*)}]
	\item {Queries ($n_c$):} The total number of expensive computations performed, which directly measures the computational cost and adherence to the budget.
	\item {Efficiency:} The expected number of true discoveries per expensive computation, which captures the return on investment:
	$
	\text{Efficiency} = \E \left[ {S}/{n_c} \right],
	$
	where the ratio is defined as zero if $n_c=0$.
\end{enumerate}

\subsection{Performance with an Auxiliary Signal}\label{sec:sim1}

In this experiment, we assess performance in a scenario where the auxiliary statistic provides a direct but unquantifiable signal about the true effect. 
Additional simulations are provided in Section \ref{sec:numadd} of the Supplement.
We simulate $N = 10,000$ hypotheses, each defined by a signal strength parameter $\mu_i$. The $i$th null hypothesis is $H_{0,i}: \mu_i = 0$.
The signal strengths $\{\mu_i\}_{i=1}^N$ are generated independently from a two-component mixture model to create a fraction $\pi$ of non-nulls:
$$
\mu_i \overset{\text{i.i.d.}}{\sim} (1-\pi)\delta_0 + \pi |\mathcal{N}(0, \tau^2)|.
$$
Let $\tau^2 = 2\log N$. 
From each primary observation $Z_i \sim \mathcal{N}(\mu_i, 1)$, we construct a corresponding gold-standard $e$-value and $p$-value:
$
E_i = \exp\left(\lambda Z_i - {\lambda^2/2}\right)  \text{ and }  P_i = 1 - \Phi(Z_i),
$
where $\lambda = \{\log(N/\alpha)\}^{1/2}$ as recommended in \cite{xu2025active} and $\Phi$ is the standard normal CDF. The corresponding auxiliary statistics, which encode the signal strength $\mu_i$, are generated as:
$
E_i^a \sim \mathrm{Poisson}(1 + \mu_i) \text{ and } P_i^a \sim \mathrm{Beta}(1, 1+\mu_i).
$
The Poisson statistic serves as a direct signal for the $e$-value, while the Beta statistic provides an inverse signal for the $p$-value.

\begin{figure}[ht!]
	\centering
		\includegraphics[width=\linewidth]{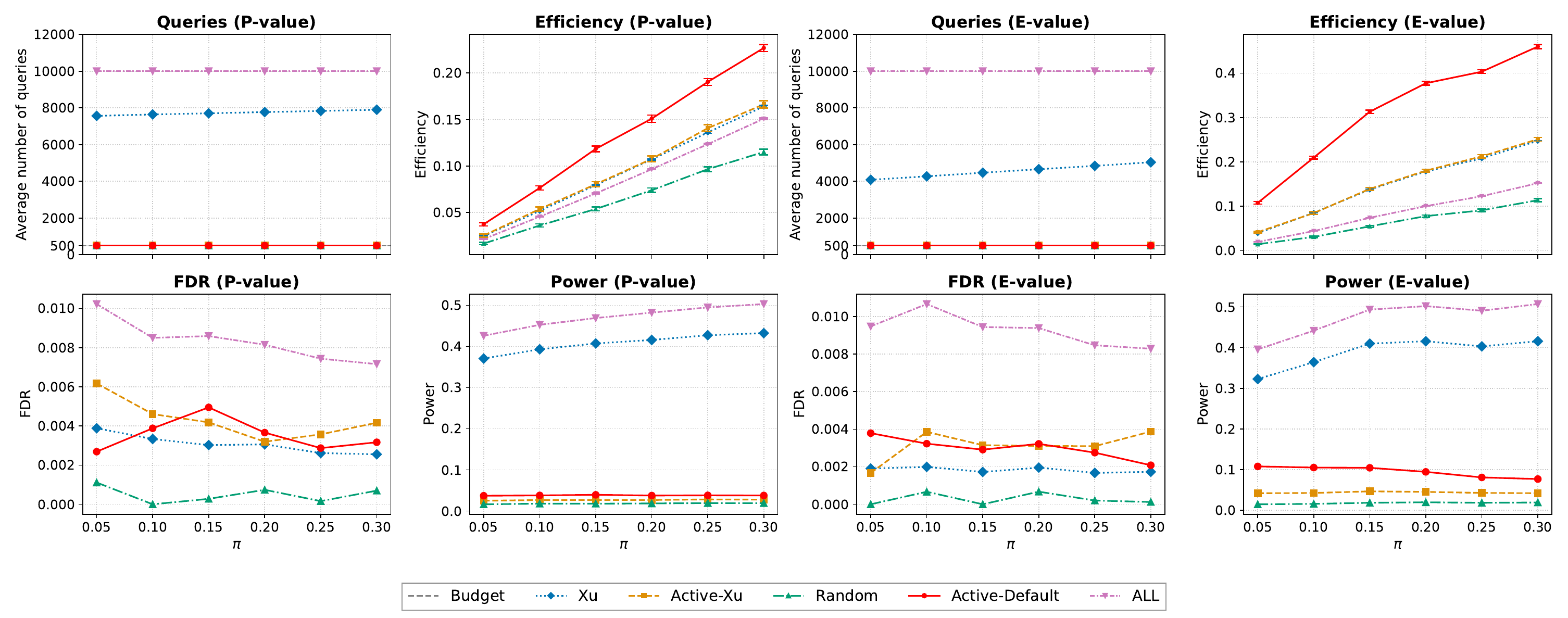}
	\caption{Performance comparison as a function of $\pi$ with a budget of $n_b=500$. All methods successfully control the FDR at $\alpha=0.1$. \texttt{Active-Default} achieves the highest efficiency. Error bars on efficiency show pointwise 95\% Monte Carlo confidence intervals for the mean over 100 repetitions.}
	\label{fig:pi_poi}
\end{figure}

\begin{figure}[ht!]
	\centering
		\includegraphics[width=\linewidth]{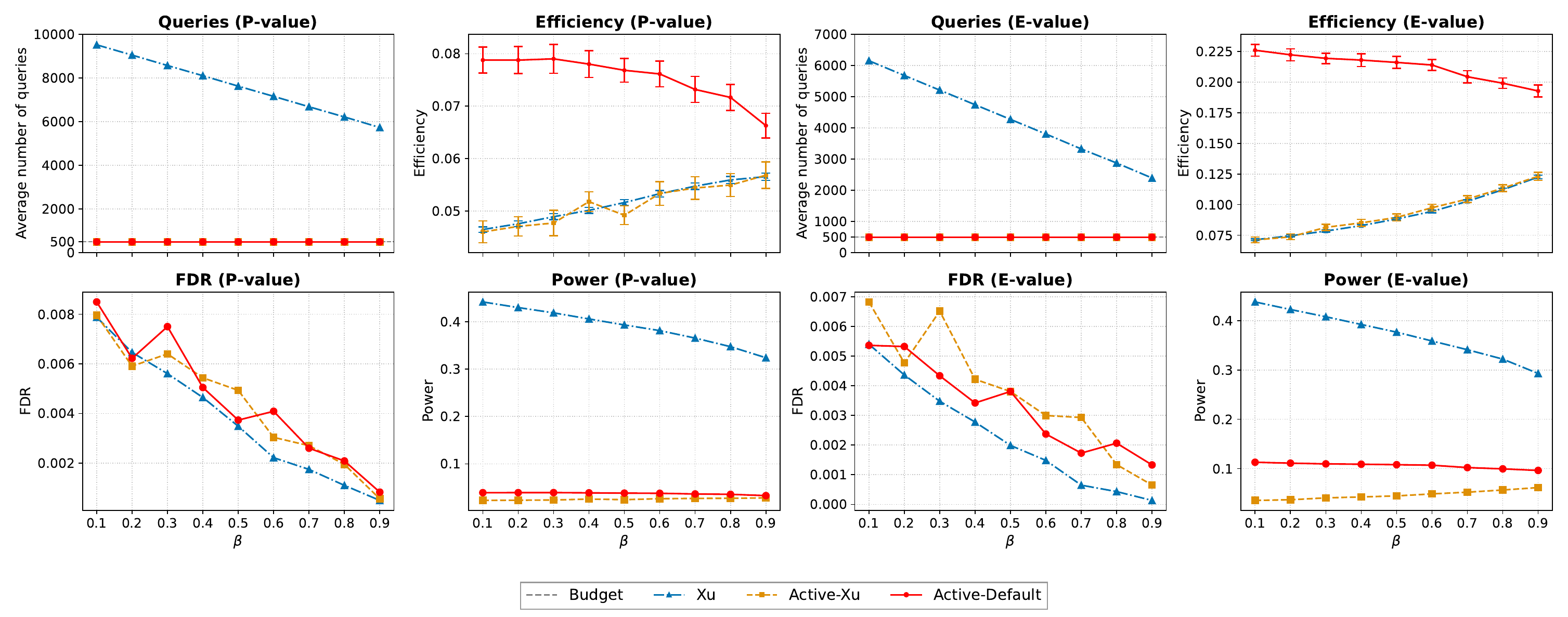}
	\caption{Performance comparison as a function of the hyperparameter $\beta$ with a budget of $n_b=500$. The choice of $\beta$ influences efficiency, but no single value dominates.  Error bars on efficiency show pointwise 95\% Monte Carlo confidence intervals for the mean over 100 repetitions.}
	\label{fig:beta_poi}
\end{figure}

We conduct two analyses based on this setup, with a computational budget of $n_b = 500$. First, to assess performance as the proportion of non-null hypotheses varies, we vary the non-null proportion $\pi$ from 0.05 to 0.3 while holding $\beta=0.5$ fixed. Second, to examine the influence of the $\beta$, we vary it from 0.1 to 0.9 while keeping $\pi=0.1$ fixed. The target FDR level is 0.1.
Given that the statistics are generated independently for each hypothesis, we employ the active $p$-value construction designed for the independent case as in \eqref{eq:p-value-indep}. The results for each analysis, averaged over 100 simulations, are presented in Figures~\ref{fig:pi_poi} and \ref{fig:beta_poi}, respectively.


The results in {Fig.}~\ref{fig:pi_poi} clearly demonstrate the practical advantages of our globally budgeted framework.
First, the empirical FDR remains below the nominal level for all procedures. The Queries panel confirms that \texttt{Active-Default}, \texttt{Active-Xu}, and \texttt{Random} adhere perfectly to the $n_b=500$ budget. In contrast, \texttt{Xu}'s query count grows with $\pi$, exceeding the prescribed budget in both settings.

The central finding lies in the interplay between Power and Efficiency. While the unconstrained \texttt{Xu} and \texttt{ALL} methods achieve higher absolute power, they do so at an enormous computational cost. When performance is measured by efficiency, our proposed \texttt{Active-Default} is the unambiguous winner. Its efficiency grows with $\pi$, indicating that its allocation strategy becomes increasingly effective as the density of true signals increases.

Furthermore, the comparison between \texttt{Xu} and \texttt{Active-Xu} is particularly revealing. By embedding the \texttt{Xu} decision logic within our global budget framework, \texttt{Active-Xu} achieves nearly identical efficiency to its unconstrained counterpart while strictly respecting the budget. This demonstrates the modularity and effectiveness of our allocation scheme. Overall, in a resource-constrained setting where return on investment is paramount, \texttt{Active-Default} provides the best performance.

In {Fig.}~\ref{fig:beta_poi}, we examine the impact of varying the hyperparameter $\beta$ from 0.1 to 0.9 while holding $\pi = 0.1$ fixed. We observe that as $\beta$ increases, the efficiency of \texttt{Active-Default} decreases, while the efficiency of \texttt{Xu} and \texttt{Active-Xu} increases. However, as we discussed in Section~\ref{sec:hbeta}, there is no uniformly optimal choice of $\beta$ that dominates across all data-generating mechanisms. The relative performance of different methods depends on the specific characteristics of the problem. Consequently, in practice, we recommend adopting the default choice of $\beta = 0.5$, which provides a balanced compromise across a wide range of scenarios. 

\section{Real-Data Analysis }\label{sec:real_data}
\subsection{Myocardial Infarction GWAS}
To demonstrate the practical utility of our framework, we apply it to a common challenge in genomics: leveraging public summary statistics from a GWAS of a related phenotype to guide discovery in a target phenotype under a computational budget. The same framework naturally extends to the same disease across distinct populations or regions, leveraging public GWAS from one group to guide discovery in another (e.g., East Asians vs. Europeans). 

Our goal is to identify single-nucleotide polymorphisms (SNPs) associated with myocardial infarction (MI). We use summary statistics from a large GWAS on hypertension (HTN) as inexpensive, auxiliary information. This scenario models a workflow where a research group might repurpose public data to prioritize which SNPs to {analyse} in their own cohort, thereby saving resources.

We obtained publicly available GWAS summary statistics from the OpenGWAS database. The target phenotype is MI (study ID: `ebi-a-GCST90038610'), \url{https://opengwas.io/datasets/ebi-a-GCST90038610} and the auxiliary phenotype is HTN (study ID: `ebi-a-GCST90038604'), \url{https://opengwas.io/datasets/ebi-a-GCST90038604} \citep{donertas2021myocardial,donertas2021hypertension}.

After aligning the two studies by their SNP identifiers (rsID), we retained $N=9,567,070$ common SNPs.
For the $i$th SNP we have its $p$-value from the HTN study, denoted by $P_i^a$, and its $p$-value from the MI study, denoted by $P_i$. A crucial distinction is that $P_i^a$ is only a valid $p$-value under the null hypothesis of no association with HTN. Under our target null hypothesis (no association with MI), the distribution of $P_i^a$ is unknown. We therefore treat $\{P_i^a\}_{i=1}^N$ as a set of auxiliary statistics. The MI $p$-values $\{P_i\}_{i=1}^N$ represent the expensive ``gold-standard" evidence whose computation we aim to limit.

We apply our \texttt{Active-Default} framework to this task. Since small $p$-values are the signal of interest (an inverse signal), {we use the utility function $1/(x + \epsilon)$ with $\epsilon=10^{-8}$ as the base utility function}. Both GWAS derive from the same UK Biobank cohort \citep{donertacs2021common}. The reported analysis uses \eqref{eq:p-value-indep}, but its independence assumption has not been established here; this comparison therefore assesses empirical query efficiency without an application-level FDR guarantee.
We include \texttt{Random}, \texttt{Xu} and \texttt{Active-Xu} for comparison.

\begin{figure}[ht!]
	\centering
	\includegraphics[width=0.9\textwidth]{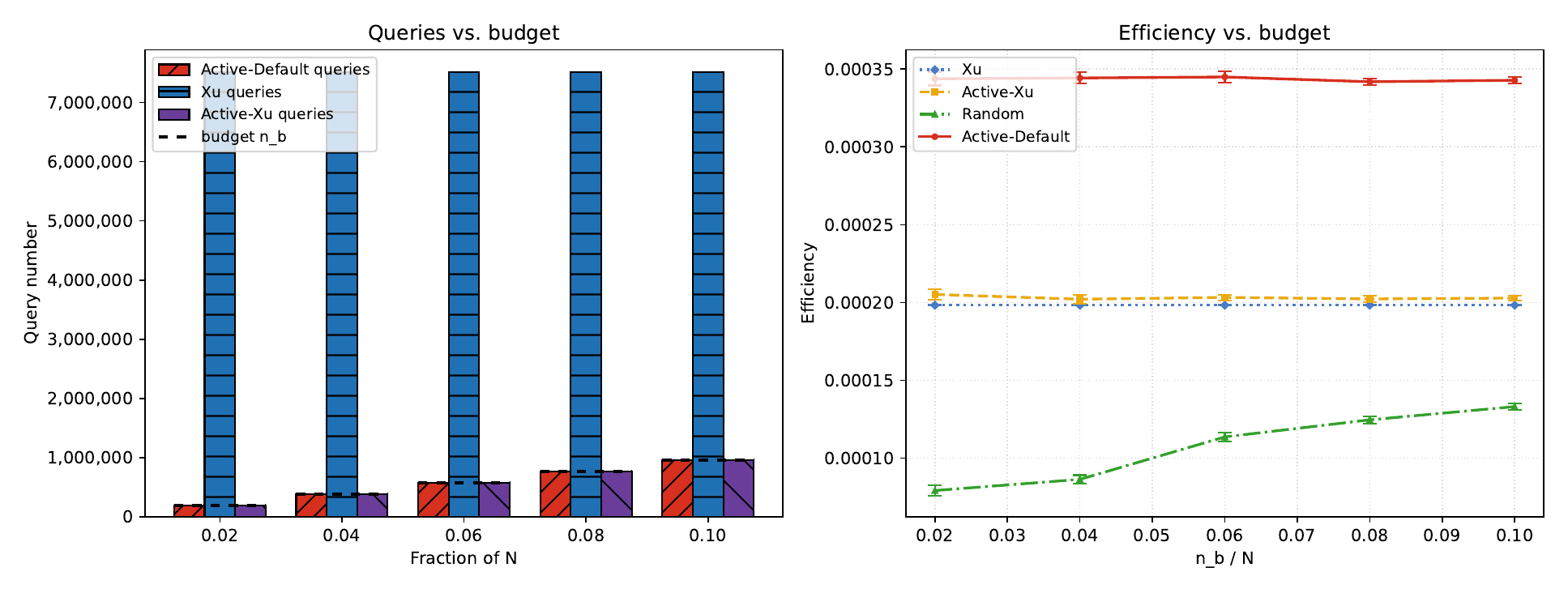} 
	\caption{Performance on the GWAS data analysis. (Left) The number of queried MI $p$-values versus the budget. (Right) The efficiency (oracle discoveries per query) versus the budget. Bars and points show means over 200 query randomizations at each budget, with the GWAS data fixed. Efficiency error bars show pointwise 95\% Monte Carlo confidence intervals for these means.} 
	\label{fig:realdata1}
\end{figure}

Since the ground truth is unknown, we establish an oracle set of discoveries to serve as a benchmark, defined as the SNPs rejected by the BY procedure at $\alpha=0.1$ on the full set of $N$ MI $p$-values. We compare the performance of our \texttt{Active-Default} method against \texttt{Random}, \texttt{Xu}, and \texttt{Active-Xu} by measuring their ability to recover these oracle discoveries. For each method, we generate active $p$-values and apply the BY procedure to identify discoveries. Performance is quantified by \textit{efficiency}, defined as the number of oracle discoveries recovered per MI $p$-value queried. We evaluate this efficiency as the budget, $n_b$, is varied as a fraction of the total number of SNPs, with the results summarized in {Fig.}~\ref{fig:realdata1}.

The left panel of {Fig.}~\ref{fig:realdata1} confirms that \texttt{Active-Default} and \texttt{Active-Xu} precisely adhere to the specified budget, while \texttt{Xu} is unable to provide a budget guarantee. The right panel demonstrates the practical benefit of our approach. At every budget level, the \texttt{Active-Default} method is substantially more efficient than \texttt{Random} and \texttt{Xu}. This indicates that the HTN summary statistics, while not directly valid for inference, provide valuable information for prioritizing the analysis of MI associations, and our framework successfully exploits this information to maximize the return on the computational budget.

To provide external validation, we examine one of the top signals prioritized and discovered by our method, rs1333047. This SNP is located in the 9p21.3 locus, which is one of the most well-established and replicated risk loci for cardiovascular disease. A recent meta-analysis confirmed its strong association with coronary artery disease and MI \citep{PAQUETTE2017406}, suggesting that testing for this variant could be an important modifier of cardiovascular risk assessment. 

\subsection{Myocardial Infarction Complications}\label{sec:real2}
Our second real-data application also addresses myocardial infarction. We utilize a dataset from \cite{Golovenkin2020} that contains clinical information for 1,700 patients. The primary objective is to predict in-hospital mortality. The original dataset features a multi-class outcome with eight categories: survival and seven distinct causes of death. For our analysis, we simplify this into a testing problem: survival versus mortality, irrespective of the specific cause.

The dataset is partitioned into a training set (800 patients with 672 alive and 128 dead), a calibration set (400 alive patients), and a test set (500 patients with 357 alive and 143 dead). We frame the problem as a multiple hypothesis testing task, where each null hypothesis $H_{0,i}$ posits that patient $i$ will survive.

The exact $p$-value, $P_i$, is constructed using the conformal inference framework \citep{bates2023testing}. To implement this, we partition the data into training, calibration ($n=400$), and test sets. We first train a random forest classifier on the full-feature training data to define a conformity score function, $\hat{s}(\cdot)$, where $\hat{s}(x)$ is a patient's predicted probability of survival. The conformal $p$-value for each test patient $i$ is then calculated by ranking their score against the scores from the calibration set $\mathcal{D}^{\mathrm{cal}}$:
\[
P_i = \frac{1+\left|\left\{j\in\mathcal{D}^{\mathrm{cal}}:\hat{s}\left(X_{j}\right)\leq\hat{s}(X_i)\right\}\right|}{n+1}.
\]

However, the computation of this exact $p$-value, $P_i$, relies on a full feature set, which includes some variables that are costly to acquire. A key example is `ZSN\_A', a feature that indicates the presence of chronic heart failure (HF). A definitive diagnosis of HF requires a comprehensive clinical assessment, including symptoms, physical signs, chest X-rays, and echocardiography. The latter, in particular, is an expensive imaging procedure requiring specialized equipment and trained personnel. In our experimental setup, we treat `ZSN\_A' as an expensive feature subject to a budget constraint. To perform hypothesis testing under this budget, it is necessary to construct an auxiliary statistic, $P_i^a$, without access to `ZSN\_A'.

To construct the auxiliary statistic $P_i^a$, we use a large language model (LLM), Gemini 3.1 Pro, to impute the costly feature.
We prompt the LLM to impute `ZSN\_A' for each patient based on their other clinical data. Using this imputed dataset, we then compute a proxy conformal $p$-value, $P_i^a$, to serve as our auxiliary statistic. The complete prompt, the full conversation history, and the attached dataset are available at  \url{https://gemini.google.com/share/4922ddaad736}.



\begin{figure}[ht!]
	\centering
	\includegraphics[width=\textwidth]{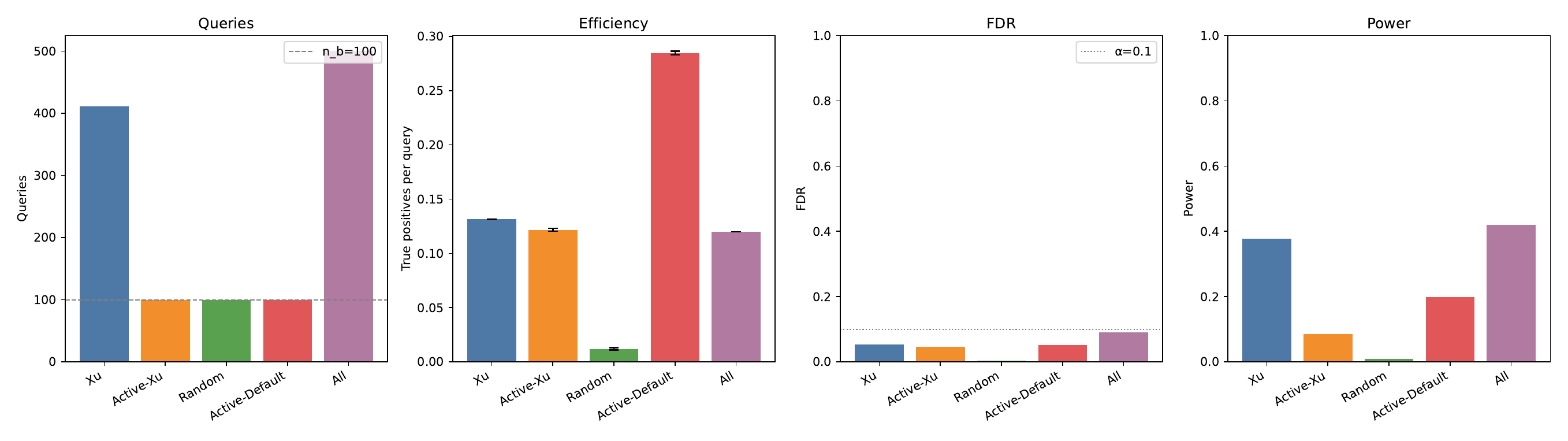} 
	\caption{Performance on the MI Complications data analysis. Results are means over 5,000 query randomizations with the fitted model, data split and LLM imputations fixed. Error bars on efficiency show pointwise 95\% Monte Carlo confidence intervals for the mean. \texttt{Active-Default} achieves the highest mean efficiency while adhering to the budget.}
	\label{fig:realdata2}
\end{figure}

We then apply our \texttt{Active-Default} framework with a budget of $n_b=100$ and a significance level of $\alpha = 0.1$ and compare it with \texttt{Random}, \texttt{Xu}, \texttt{ALL}, and \texttt{Active-Xu} baselines. As with the GWAS analysis, we treat the auxiliary statistics as inverse signals and use the base utility function $1/(x+\epsilon)$. The other settings are the same as in the previous subsection. Discoveries are identified using the BH procedure. The conditional validity and PRDS assumptions of Theorem~\ref{thm:active_bh_conditional} have not been established for this application; the reported mean FDPs therefore provide an empirical comparison rather than a guarantee of FDR control. The results are presented in {Fig.}~\ref{fig:realdata2}.
The mean FDPs are below $\alpha = 0.1$, while the procedures exhibit marked differences in budget adherence. The budgeted methods (\texttt{Active-Default}, \texttt{Active-Xu}, and \texttt{Random}) precisely respect the $n_b = 100$ query limit, whereas the unconstrained \texttt{Xu} and \texttt{ALL} methods require significantly more computations. This highlights a clear trade-off between statistical power and computational cost. Although  \texttt{ALL} and \texttt{Xu} achieve higher absolute power, our proposed \texttt{Active-Default} demonstrates the highest efficiency, delivering the greatest number of discoveries per query. The comparison between \texttt{Xu} and \texttt{Active-Xu} also illustrates the computational savings obtained by enforcing the query budget. These findings collectively demonstrate that in resource-constrained settings where efficiency is paramount, our \texttt{Active-Default} framework provides a powerful and principled solution.

\section{Discussion}\label{discussion}

The practical implementation of our framework is guided by the choice of utility functions $\{u_i\}$ and the hyperparameter $\beta$. While our admissibility results (Propositions~\ref{prop:no-optimal-h} and \ref{prop:no-optimal-beta}) show that no universally optimal choice exists, we have provided guidance that yields effective performance in practice. A natural direction for future work is to learn utility functions from data, for example by using a held-out calibration dataset to tune $u_i(\cdot)$ to maximize a downstream objective such as the number of discoveries. Further work could incorporate graph structure into budget allocation and extend the framework to online testing, where resources must be allocated as hypotheses arrive.

\section*{Declaration of the use of generative AI and AI-assisted technologies}

During the preparation of this work, the authors used Gemini 3.1 Pro in order to polish the language and perform professional proofreading to improve the readability of the manuscript. Gemini 3.1 Pro was also used for the feature imputation in Section~\ref*{sec:real2}. After using AI, the authors reviewed and edited the content as needed and take full responsibility for the content of the publication.

\section*{Supplementary material}
The Supplementary Material includes dominance comparisons, multivariate admissibility results, additional simulations, technical proofs and counterexamples.
\bibliographystyle{plainnat}
\bibliography{reference}

\newpage

\begin{center}
	 \begin{spacing}{2}
	{\huge Supplementary Material to ``Active Hypothesis Testing under Computational Budgets''}
	 \end{spacing}
\end{center}
This supplement contains the dominance results of direct construction in Section \ref{sec:appendix_dominance}, admissibility in multivariate setting in Section~\ref{sec:admiss-multi}, additional numerical experiments in Section~\ref{sec:numadd},  technical proofs in Section \ref{sec:proof}, relevant counterexamples in Sections \ref{sec:counterexample} and \ref{sec:counterexample-decomposition}, {{and a detailed comparison with the framework of \citet{xu2025active} in Section~\ref{sec:comparison_xu}.}}
\numberwithin{equation}{section}
\numberwithin{figure}{section}
\numberwithin{theorem}{section}
\numberwithin{definition}{section}
\numberwithin{assumption}{section}
\numberwithin{proposition}{section}
\numberwithin{remark}{section}

\appendix

\section{Dominance of Direct Construction for Active Statistics}
\label{sec:appendix_dominance}
\subsection{{The Connection Between $p$-values and $e$-values via Calibrators}}

In the main text, we present separate constructions for active $p$-values and active $e$-values. This appendix compares direct and indirect constructions and identifies the conditions under which the direct construction dominates. To formalize this comparison, we first establish the mathematical tools used for such conversions.

\label{sec:appendix_calibrators}

To define an indirect construction path (e.g., constructing an active $p$-value from an active $e$-value), we require a principled method for converting between statistic types. This is the role of calibrators.

A {$p$-to-$e$ calibrator} is a decreasing function \(f:[0,\infty)\to[0,\infty]\) that is zero on \((1,\infty)\) such that for any valid $p$-value $P$, the transformed variable $f(P)$ is a valid $e$-value \citep{vovk2021values}. {An example is \(f(p)=-\log p\) for \(0<p\le1\), extended by \(f(0)=\infty\) and \(f(p)=0\) for \(p>1\).} These calibrators share a fundamental property, formalized in the following lemma.

\begin{lemma}
	\label{lem:calibrator_bound}
	A $p$-to-$e$ calibrator $f$ must satisfy the inequality \(f(x) \cdot x \leq 1\) for all $x \in {(0,1]}$.
\end{lemma}

This lemma is used in the comparisons below.

Conversely, conversion from $e$-values to $p$-values is more constrained. {The reciprocal $1/E$ satisfies the super-uniformity inequality, and its truncation $\min\{1,1/E\}$ is the only admissible $e$-to-$p$ calibrator} \citep{vovk2021values}. {We specify below whether truncation is applied in each comparison.}
Equipped with these definitions, we can now formally compare the direct and indirect construction methods for both active $p$-values and active $e$-values.

\subsection{{Comparison of Direct and Indirect Active $p$-value Constructions}}
\label{sec:appendix_dominance_p}

{
We compare the direct construction with the indirect construction that first converts $p$-values into $e$-values. Let $P$ be a valid $p$-value and $P^a\in[0,1]$ an auxiliary statistic. Fix $\beta\in(0,1)$ and a $p$-to-$e$ calibrator $f$. Let $h_e$ be a measurable control function taking values in $[0,1]$, defined on the range of $f$, including $\infty$ if needed. Set $h_p=h_e\circ f$, so that both constructions query on the same event $U<h_p(P^a)$, using the same $U\sim\mathrm{Uniform}(0,1)$ independent of $(P,P^a)$. All final $p$-values are compared after truncation at one.

\par\smallskip\noindent\textit{Indirect Construction (via $e$-values)}\enspace
Set $E=f(P)$ and $E^a=f(P^a)$. The intermediate active $e$-value is
\[
E_{\mathrm{indirect}}^{\mathrm{active}}=
\begin{cases}
\dfrac{\beta}{1-h_e(E^a)} & \text{if }U\ge h_e(E^a),\\[1em]
\dfrac{1-\beta}{h_e(E^a)}E & \text{if }U<h_e(E^a).
\end{cases}
\]
Taking its reciprocal gives $P_{\mathrm{indirect}}^{\mathrm{active}}=1/E_{\mathrm{indirect}}^{\mathrm{active}}$, with $1/0=\infty$ and $1/\infty=0$. Since $h_e(f(P^a))=h_p(P^a)$, this is
\[
P_{\mathrm{indirect}}^{\mathrm{active}}=
\begin{cases}
\dfrac{1-h_p(P^a)}{\beta} & \text{if }U\ge h_p(P^a),\\[1em]
\dfrac{h_p(P^a)}{1-\beta}\dfrac{1}{f(P)} & \text{if }U<h_p(P^a).
\end{cases}
\]

\par\smallskip\noindent\textit{Direct Construction}\enspace
\label{sec:appendix_direct_p}
Under independence of $P$ and $P^a$, the direct construction in \eqref{eq:p-value-indep}, with the same $h_p$ and $\beta$, is
\[
P_{\mathrm{direct}}^{\mathrm{active}}=
\begin{cases}
\dfrac{1-h_p(P^a)}{\beta} & \text{if }U\ge h_p(P^a),\\[1em]
\dfrac{h_p(P^a)}{1-\beta}P & \text{if }U<h_p(P^a).
\end{cases}
\]

\par\smallskip\noindent\textit{Comparison under Independence}\enspace
\label{sec:appendix_dominance_p_proof}
On the nonquery branch, the two outputs are identical. On the query branch, $h_p(P^a)>0$. For $P>0$, Lemma~\ref{lem:calibrator_bound} gives $Pf(P)\le1$, hence $P\le1/f(P)$ under the reciprocal conventions above. The latter inequality also holds when $P=0$. Therefore,
\[
P_{\mathrm{direct}}^{\mathrm{active}}
=\frac{h_p(P^a)}{1-\beta}P
\le\frac{h_p(P^a)}{1-\beta}\frac{1}{f(P)}
=P_{\mathrm{indirect}}^{\mathrm{active}}.
\]
Truncation preserves the inequality, so in both branches
\[
\min\{1,P_{\mathrm{direct}}^{\mathrm{active}}\}
\le\min\{1,P_{\mathrm{indirect}}^{\mathrm{active}}\}
\quad\text{almost surely}.
\]
Thus the direct construction is at least as powerful as this indirect construction when the same rejection threshold is used. The comparison after truncation is strict on the query branch whenever $P<1/f(P)$ and $h_p(P^a)P/(1-\beta)<1$.

\par\smallskip\noindent\textit{Comparison under General Dependence}\enspace
For the direct construction in \eqref{eq:p-value-general}, write $M=\sup_x h_p(x)$. Its query output is $MP/(1-\beta)$, while the indirect query output remains $h_p(P^a)/\{(1-\beta)f(P)\}$. These quantities do not have a uniform ordering. For example, take $f(p)=-\log p$ on $(0,1]$, $\beta=1/2$, $P^a\equiv1$, $h_p(1)=1/10$ and $h_p(x)=1$ for $0\le x<1$. This choice is compatible with $h_p=h_e\circ f$ by taking $h_e(0)=1/10$ and $h_e(x)=1$ for $x>0$, including $x=\infty$. Then $M=1$, and on $U<1/10$ the two outputs are $2P$ and $1/\{5(-\log P)\}$. At $P=e^{-5}$ the first is smaller, since $2e^{-5}<1/25$; at $P=e^{-1}$ the second is smaller, since $1/5<2/e$. Both comparisons occur below one and persist on intervals of positive probability when $P\sim\mathrm{Uniform}(0,1)$. These valid inputs are permitted in the general-dependence class, so neither construction uniformly dominates the other in that class.
}

\subsection{Dominance of the Direct Active $e$-value Construction}
\label{sec:appendix_dominance_e}

{We compare the direct active $e$-value with the indirect construction below, which uses untruncated reciprocals.} Let $E$ be the exact $e$-value and $E^a$ be the auxiliary $e$-value. {Fix $\beta\in(0,1)$ and a measurable control function $h_p:[0,\infty]\to[0,1]$. Use the same $U\sim\mathrm{Uniform}(0,1)$, independent of $(E,E^a)$, in both constructions, and set $h_e(x)=h_p(1/x)$, with $1/0=\infty$. Extend the calibrator $f$ by $f(\infty)=0$.}

\par\smallskip\noindent{\textit{Indirect Construction (via $p$-values)}}\enspace 
\label{sec:appendix_indirect_e}

The indirect construction for an active $e$-value proceeds as follows:
\begin{enumerate}
	\item {Conversion to $p$-values:} Transform the $e$-values using the reciprocal calibrator: $P = 1/E$ and $P^a = 1/E^a$.
	\item {Active $p$-value construction:} Given a control function $h_p$ and a $\beta$, construct the active $p$-value $P_{\mathrm{indirect}}^{\mathrm{active}}$ using the formulation from the main text (e.g., Equation \eqref{eq:p-value-indep} or \eqref{eq:p-value-general}). This yields:
	$$
	P_{\mathrm{indirect}}^{\mathrm{active}} =
	\begin{cases}
		\dfrac{1 - h_p(P^a)}{\beta} & \text{if } U \geq h_p(P^a) \\
		b(P^a)P & \text{if } U < h_p(P^a)
	\end{cases}
	$$
	{On the query branch, take $b(q)=h_p(q)/(1-\beta)$ when $E\perp E^a$, or $b(q)=\sup_{y\in[0,\infty]}h_p(y)/(1-\beta)$ under arbitrary dependence. These are the two constructions in Theorem~\ref{thm:active_p_admissible}, with the auxiliary domain extended to $[0,\infty]$. Their validity arguments apply unchanged to the untruncated $P$, since $\pr(P\le s)=\pr(E\ge1/s)\le s$ for $s\in(0,1]$.}
	\item {Conversion back to $e$-value:} Apply a $p$-to-$e$ calibrator $f$ to obtain the final active $e$-value: $E_{\mathrm{indirect}}^{\mathrm{active}} = f(P_{\mathrm{indirect}}^{\mathrm{active}})$.
\end{enumerate}

\par\smallskip\noindent{\textit{Direct Construction}}\enspace 
\label{sec:appendix_direct_e}

The direct construction of an active $e$-value, $E_{\mathrm{direct}}^{\mathrm{active}}$, for a control function $h_e$ (related to $h_p$ via $h_e(x) = h_p(1/x)$) is given by:
$$
E_{\mathrm{direct}}^{\mathrm{active}} =
\begin{cases}
	\dfrac{\beta}{1 - h_e(E^a)} & \text{if } U \geq h_e(E^a) \\
	\dfrac{1 - \beta}{h_e(E^a)}E & \text{if } U < h_e(E^a)
\end{cases}
$$

\par\smallskip\noindent{\textit{Proof of Dominance}}\enspace 
\label{sec:appendix_dominance_e_proof}

We now show that $E_{\mathrm{direct}}^{\mathrm{active}} \geq E_{\mathrm{indirect}}^{\mathrm{active}}$ {for the specified indirect construction.} We {analyse} the two branches of the construction separately.

If $U \geq h_e(E^a)$, the corresponding indirect $p$-value is $P_{\mathrm{indirect}}^{\mathrm{active}} = {(1 - h_p(P^a))/(\beta)}$. Substituting $P^a = 1/E^a$ and $h_p(x) = h_e(1/x)$, this becomes ${(1 - h_e(E^a))/(\beta)}$. The final indirect $e$-value is therefore:
$$
E_{\mathrm{indirect}}^{\mathrm{active}} = f\left(\frac{1 - h_e(E^a)}{\beta}\right).
$$
{On this branch, $(1-h_e(E^a))/\beta>0$ almost surely. Lemma~\ref{lem:calibrator_bound} gives $f(x)\le1/x$ for $0<x\le1$, and the same bound holds for $x>1$ because $f(x)=0$. Applying it gives:}
$$
E_{\mathrm{indirect}}^{\mathrm{active}} \leq \frac{\beta}{1 - h_e(E^a)} = E_{\mathrm{direct}}^{\mathrm{active}}.
$$
Thus, the direct construction dominates in this branch.

If $U < h_e(E^a)$, the indirect $p$-value is $P_{\mathrm{indirect}}^{\mathrm{active}} = b(P^a)P = b(1/E^a)/E$. The final indirect $e$-value is:
$$
E_{\mathrm{indirect}}^{\mathrm{active}} = f\left(\frac{b(1/E^a)}{E}\right).
$$
{On the query branch, $h_e(E^a)>0$ and $b(1/E^a)>0$. If $E=0$, the indirect output is $f(\infty)=0$, equal to the direct output. For $E>0$, the same calibrator bound gives:}
$$
E_{\mathrm{indirect}}^{\mathrm{active}} \leq \frac{E}{b(1/E^a)}.
$$
{Both specified choices of $b$ satisfy $b(P^a)\ge h_p(P^a)/(1-\beta)$ on the query branch. Consequently, $1/b(P^a)\le(1-\beta)/h_p(P^a)$, and}
$$
E_{\mathrm{indirect}}^{\mathrm{active}} \leq E \cdot \frac{1-\beta}{h_p(1/E^a)} = E \cdot \frac{1-\beta}{h_e(E^a)} = E_{\mathrm{direct}}^{\mathrm{active}}.
$$
The direct construction also dominates in this second branch.

\par\smallskip\noindent{\textit{Conclusion.}}\enspace  {For this indirect construction using untruncated reciprocals, the direct active $e$-value is point-wise no smaller, and hence has at least as much power at any common rejection threshold.}


\section{Admissibility in Multivariate Setting}\label{sec:admiss-multi}
In Section~\ref{sec:hbeta} of the main text, we established the admissibility of active statistics for a single hypothesis, focusing on the scalar control function $h(\cdot)$ and hyperparameter $\beta$. However, under the global budget framework, the decision probabilities for individual hypotheses are coupled through the budget constraint. Consequently, the control function becomes a multivariate vector mapping the full set of auxiliary statistics $\mathbf{X}^a = (X^a_1, \dots, X^a_N)$ to a vector of probabilities. This section extends the concept of admissibility to this multivariate, budget-constrained setting. We formalize domination via component-wise vector comparisons and prove that no feasible allocation strategy uniformly dominates another.

To satisfy the global budget constraint, the vector of control functions $\mathbf{h} = (h_1, \dots, h_N)$ must satisfy
\[
\sum_{i=1}^N h_i(\mathbf{X}^a) = n_b
\]
for any realization of $\mathbf{X}^a$. We denote by $\mathcal{H}$ the set of all measurable control function vectors $\mathbf{h}: \mathcal{X}^N \to [0, 1]^N$ that satisfy the above equality. The concept of domination extends naturally to the multivariate case by comparing vectors of active statistics.

\begin{definition}[Multivariate Domination and Admissibility]
	\label{def:multi_domination}
	Let $\mathbf{X}^{\text{active}}_{\mathbf{h},\boldsymbol{\beta}}$ denote the vector of active statistics induced by a control vector $\mathbf{h} \in \mathcal{H}$ and a hyperparameter vector $\boldsymbol{\beta} \in (0, 1)^N$.
	\begin{enumerate}
		\item {For $p$-values:} The vector $\mathbf{X}^{\text{active}}_{\tilde{\mathbf{h}},\tilde{\boldsymbol{\beta}}}$ {dominates} $\mathbf{X}^{\text{active}}_{\mathbf{h},\boldsymbol{\beta}}$ if, for any valid input, the component-wise inequality
		\[
		\min\{\mathbf{1}, \mathbf{X}^{\text{active}}_{\tilde{\mathbf{h}},\tilde{\boldsymbol{\beta}}}\} \le \min\{\mathbf{1}, \mathbf{X}^{\text{active}}_{\mathbf{h},\boldsymbol{\beta}}\}
		\]
		holds almost surely, and there exists at least one valid input distribution such that, with positive probability, the inequality is strict for at least one component.
		
		\item {For $e$-values:} The vector $\mathbf{X}^{\text{active}}_{\tilde{\mathbf{h}},\tilde{\boldsymbol{\beta}}}$ {dominates} $\mathbf{X}^{\text{active}}_{\mathbf{h},\boldsymbol{\beta}}$ if, for any valid input, the component-wise inequality
		\[
		\mathbf{X}^{\text{active}}_{\tilde{\mathbf{h}},\tilde{\boldsymbol{\beta}}} \ge \mathbf{X}^{\text{active}}_{\mathbf{h},\boldsymbol{\beta}}
		\]
		holds almost surely, and there exists at least one valid input distribution such that, with positive probability, the inequality is strict for at least one component.
	\end{enumerate}
	A vector of active statistics is {admissible} if it is not dominated by any other vector generated by a valid pair $(\tilde{\mathbf{h}}, \tilde{\boldsymbol{\beta}})$.
\end{definition}

The following propositions compare allocations or hyperparameters within the same active construction, using common input statistics and the sampling rule in Algorithm~\ref{alg:unified_framework}, driven by the same $U$ independent of the input statistics. 

\begin{proposition}[Admissibility of the Allocation Strategy]
	\label{prop:multi}
	Fix $\boldsymbol{\beta}\in(0,1)^N$. Admissibility below is with respect to changes in $\mathbf h$ within $\mathcal H$.
	\begin{enumerate}
		\item For active $e$-values, every $\mathbf h\in\mathcal H$ is admissible.
		\item For active $p$-values under either independence or general dependence, every $\mathbf h\in\mathcal H$ is admissible.
	\end{enumerate}
\end{proposition}

The fixed total budget is essential here: increasing some query probabilities requires decreasing others, so the scalar restrictions on $h$ in Proposition~\ref{prop:no-optimal-h} are not needed.

\begin{proposition}[Admissibility of the Hyperparameters]
	\label{prop:multi-beta-adm}
	Fix $\mathbf h\in\mathcal H$ such that each $h_i$ is not identically $0$ or $1$. Admissibility below is with respect to changes in $\boldsymbol\beta\in(0,1)^N$ with $\mathbf h$ fixed.
	\begin{enumerate}
		\item For active $e$-values, every $\boldsymbol\beta\in(0,1)^N$ is admissible.
		\item For active $p$-values under either independence or general dependence, $\boldsymbol\beta\in(0,1)^N$ is admissible if and only if, for every $i$, there is an auxiliary vector $\mathbf x\in[0,1]^N$ such that $1-\beta_i<h_i(\mathbf x)<1$. Thus, if $\sup\{h_i(\mathbf x):h_i(\mathbf x)<1\}=1$ for every $i$, every $\boldsymbol\beta\in(0,1)^N$ is admissible.
	\end{enumerate}
\end{proposition}


\section{Additional Numerical Experiments}\label{sec:numadd}

\subsection{Performance with a Correlated Proxy}

This simulation investigates a scenario where the gold-standard and auxiliary statistics share a deeper structural relationship {modelled} by correlation. This is representative of many real-world problems where a cheap measurement is an indirect but correlated indicator of an expensive one (e.g., gene expression levels and protein abundance).

The underlying signal structure remains the same, while the primary and auxiliary data, $Z_i$ and $Y_i$, are now drawn from a bivariate normal distribution with correlation $\rho$:
\[
\begin{bmatrix} Y_i \\ Z_i \end{bmatrix} \sim 
(1-\pi)	\mathcal{N}\left( \begin{bmatrix} 0 \\ 0 \end{bmatrix}, \begin{bmatrix} 1 & \rho \\ \rho & 1 \end{bmatrix} \right) +\pi
\mathcal{N}\left( \begin{bmatrix} \rho \mu_i \\ \mu_i \end{bmatrix}, \begin{bmatrix} 1 & \rho \\ \rho & 1 \end{bmatrix} \right).
\]
We then compute $(E_i, E_i^a)$ and $(P_i, P_i^a)$ from $(Z_i, Y_i)$ via the definitions given in \eqref{eq:true_statistics} and \eqref{eq:aux_statistics}, and the correlation $\rho$ directly controls the quality of both auxiliary channels.

We perform two analyses. First, we fix the correlation at a moderate level, $\rho=0.5$, and vary $\pi$ from 0.05 to 0.3. Second, we fix $\pi=0.1$ and vary $\rho$ from 0.2 to 0.9, assessing how well each method capitalizes on improving proxy quality. Again, we adopt the active $p$-value constructed for the general dependent case as in \eqref{eq:p-value-general}.

\begin{figure}[ht!]
	\centering
		\includegraphics[width=\linewidth]{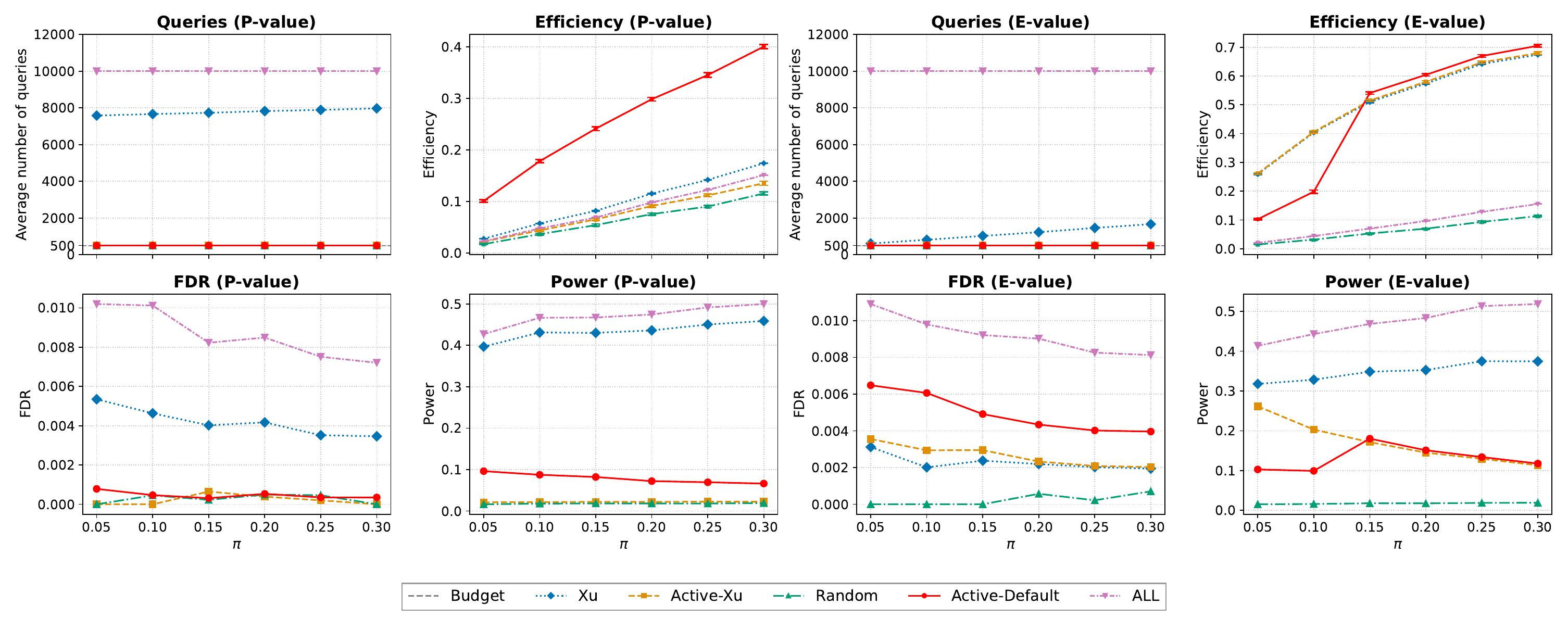}
	\caption{Performance comparison as a function of $\pi$, with a fixed $\rho=0.5$.  Error bars on efficiency show pointwise 95\% Monte Carlo confidence intervals for the mean over 100 repetitions.}
	\label{fig:pi_normal}
\end{figure}

The results of the first analysis, displayed in {Fig.}~\ref{fig:pi_normal}, confirm the robustness of our method. The performance patterns are consistent with those observed in the previous, structurally different simulations. \texttt{Active-Default} adheres to the budget while delivering the highest efficiency, with its advantage widening as the proportion of true signals grows.

\begin{figure}[ht!]
	\centering
		\includegraphics[width=\linewidth]{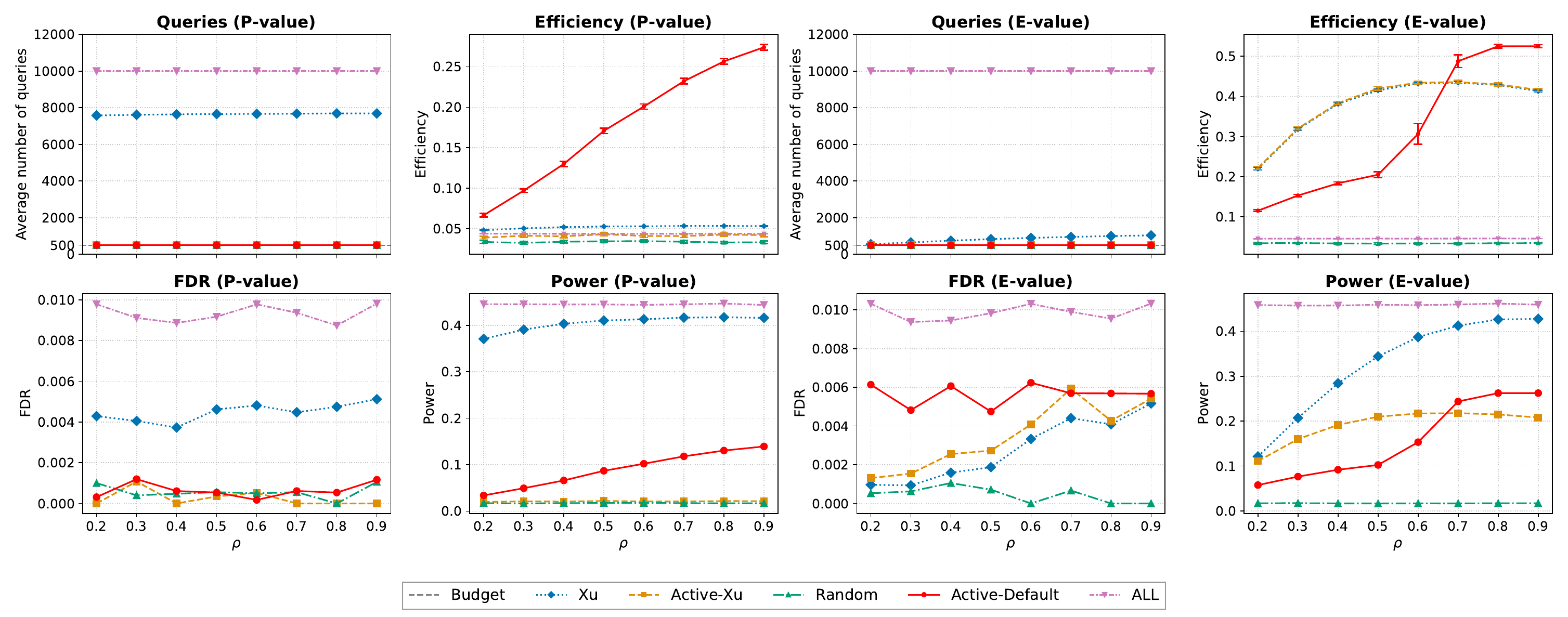}
	\caption{Performance comparison as a function of $\rho$, with a fixed $\pi=0.1$. Error bars on efficiency show pointwise 95\% Monte Carlo confidence intervals for the mean over 100 repetitions.}
	\label{fig:rho_normal}
\end{figure}

The second analysis, shown in {Fig.}~\ref{fig:rho_normal}, provides a deeper insight into the methods' {behaviour}. As $\rho$ increases, the auxiliary statistic becomes a more faithful proxy for the gold-standard statistic. This increased information quality allows all active inference methods to improve their power and efficiency. However, \texttt{Active-Default} demonstrates the most significant gains. Its efficiency curve rises more steeply than those of the other methods, highlighting its superior ability to capitalize on high-quality side information. This result shows that our framework not only works well with weak proxies but excels when strong auxiliary data are available, making it an adaptive and powerful tool for budgeted inference.

\subsection{Performance with a Noisy Proxy}

We next evaluate the methods in a ``noisy measurement" setting. This scenario models applications where the auxiliary statistic is not just a simple signal but is itself an ``$e$-value'' or a ``$p$-value'' computed from a degraded or noisy version of the primary data.
The underlying signal generation remains identical to that in Section~\ref{sec:sim1}, with hypotheses driven by a signal strength parameter $\mu_i$. The key difference lies in the construction of the auxiliary statistic. We create the noisy data $Y_i$ by 
$
Y_i = Z_i + \varepsilon_i, \text{ where } \varepsilon_i \sim \mathcal{N}(0, \sigma^2).
$
From $Z_i$ and $Y_i$ we construct both $e$-values and $p$-values in parallel:
\begin{align}
	E_i &= \exp\!\left(\lambda Z_i - \dfrac{\lambda^2}{2}\right) \qquad \text{and} \qquad P_i = 1 - \Phi(Z_i), \label{eq:true_statistics} \\
	E_i^a &= \exp\!\left(\lambda Y_i - \dfrac{\lambda^2}{2}\right) \qquad \text{and} \qquad P_i^a = 1 - \Phi(Y_i). \label{eq:aux_statistics}
\end{align}
with $\lambda = \sqrt{\log(N/\alpha)}$. Here $E_i^a$ is a direct but noisy proxy for $E_i$, and $P_i^a$ is the analogous noisy proxy for $P_i$.

We conduct two analyses within this framework. First, we fix the noise standard deviation at $\sigma=2$ and vary the non-null proportion $\pi$ from 0.05 to 0.3. Second, we fix $\pi=0.1$ and vary $\sigma$ from 1 to 5 to assess the methods' robustness to deteriorating proxy quality. Here we adopt the active $p$-value constructed for the general dependent case as in \eqref{eq:p-value-general}.

\begin{figure}[t!]
	\centering
	\includegraphics[width=\linewidth]{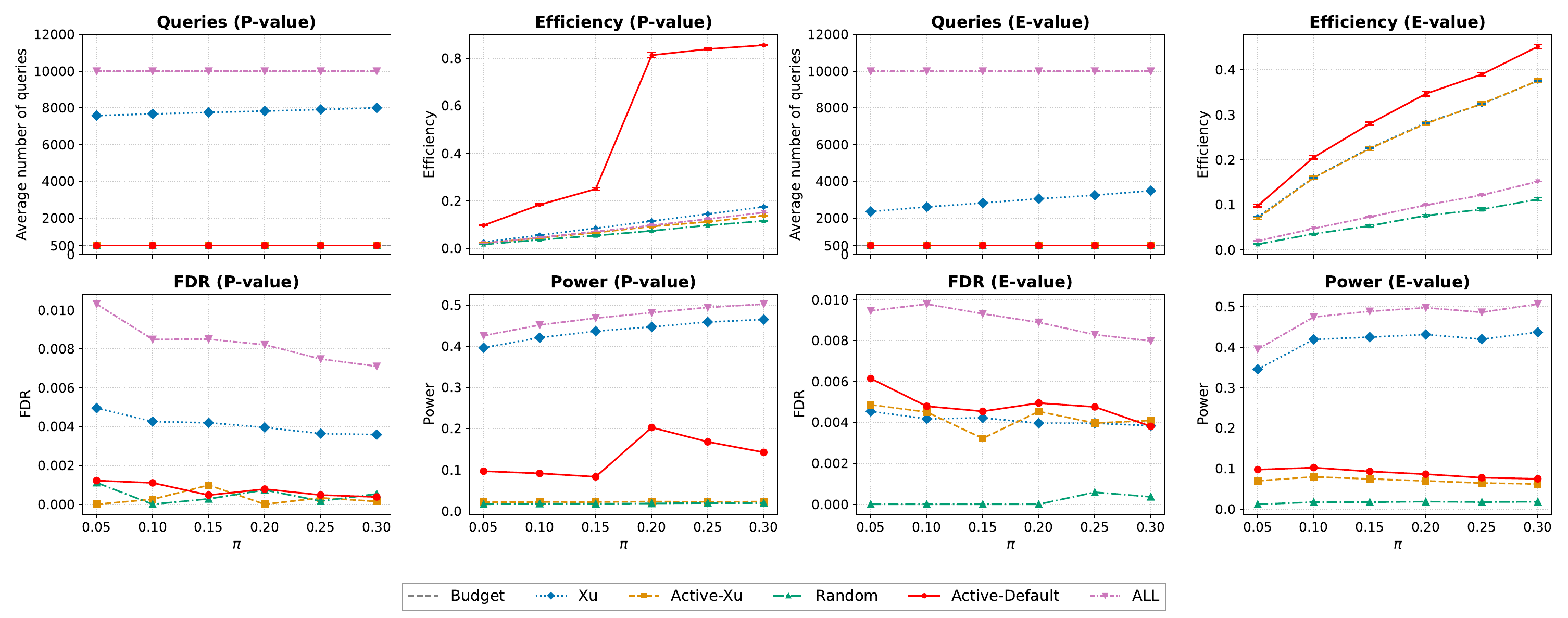}
	\caption{Performance comparison as a function of $\pi$, with a fixed $\sigma=2$.  Error bars on efficiency show pointwise 95\% Monte Carlo confidence intervals for the mean over 100 repetitions.}
	\label{fig:pi_noise}
\end{figure}

The results of the first analysis, shown in {Fig.}~\ref{fig:pi_noise}, are highly consistent with our findings from Section \ref{sec:sim1}. All methods control the FDR, and our globally budgeted approaches perfectly adhere to the $n_b=500$ query limit. The \texttt{Active-Default} method again emerges as the most efficient, with its advantage growing as the proportion of true signals increases.

\begin{figure}[t!]
	\centering
	\includegraphics[width=\linewidth]{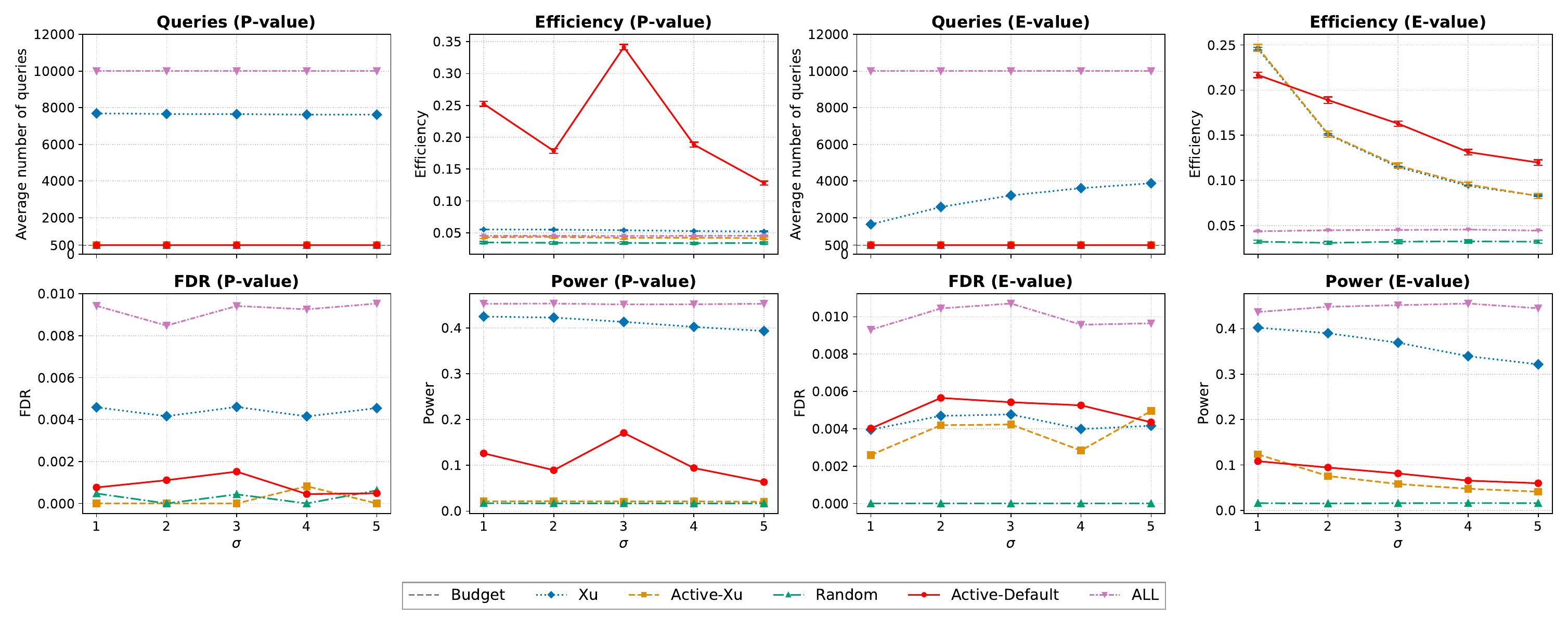}
	\caption{Performance comparison as a function of $\sigma$, with a fixed $\pi=0.1$.  Error bars on efficiency show pointwise 95\% Monte Carlo confidence intervals for the mean over 100 repetitions.}
	\label{fig:sigma_noise}
\end{figure}

The second analysis, presented in {Fig.}~\ref{fig:sigma_noise}, probes the methods' robustness. As the noise level $\sigma$ increases, the auxiliary statistic $E_i^a$ becomes a less reliable indicator of the exact $e$-value $E_i$. Consequently, the power and efficiency of \texttt{Xu}, \texttt{Active-Xu}, and \texttt{Active-Default} decline. Our \texttt{Active-Default} method maintains an efficiency advantage over the other budget-constrained methods at higher noise levels. This demonstrates that even as the quality of the auxiliary information degrades, our framework's ability to efficiently allocate a fixed budget provides a durable performance advantage.

\section{Technical Proofs}\label{sec:proof}

\subsection{Proof of Theorem \ref{thm:e-value-characterization}}

\begin{proof}
{
If statement 2 holds, independence of $U$ from $(E^a,E)$ gives
\[
\E[E^{\mathrm{active}}]
=\E[a(E^a)(1-h(E^a))]+\E[b(E^a)h(E^a)E]
\le\beta+(1-\beta)\E[E]\le1.
\]
This proves sufficiency.

For necessity, let
\[
A:=\sup_{x\ge0}a(x)(1-h(x)),\qquad
B:=\sup_{x\ge0}b(x)h(x),
\]
where either supremum may be infinite. If $A+B\le1$, then $\beta=A\in[0,1]$ satisfies statement 2. Suppose instead that $A+B>1$. By the definition of the suprema, there exist $x_1,x_2\ge0$ such that the non-negative values
\[
c_1:=a(x_1)(1-h(x_1)),\qquad c_2:=b(x_2)h(x_2)
\]
satisfy $c_1+c_2>1$. This choice does not require either supremum to be attained or finite. Choose $\epsilon\in(0,1)$ sufficiently close to one that $\epsilon c_1+c_2>1$, and define the joint law directly by
\[
(E^a,E)=
\begin{cases}
(x_1,0),&\text{with probability }\epsilon,\\
(x_2,1/(1-\epsilon)),&\text{with probability }1-\epsilon.
\end{cases}
\]
This law is well defined even if $x_1=x_2$, and $\E[E]=1$. Taking $U$ independent of this pair, we obtain
\begin{align*}
\E[E^{\mathrm{active}}]
&=\epsilon a(x_1)(1-h(x_1))
 +(1-\epsilon)a(x_2)(1-h(x_2))+b(x_2)h(x_2)\\
&\ge\epsilon c_1+c_2>1.
\end{align*}
This contradicts statement 1. Hence $A+B\le1$, and the choice $\beta=A$ proves necessity.
}
\end{proof}

\subsection{Proof of Theorem~\ref{prop:a}}

\begin{proof}
	The proof consists of two parts. First, we establish a necessary lower bound for $a(x)$ by considering a specific distribution for $P^a$—namely, a point mass at $x$. This forces $a(x)$ to satisfy a point-wise inequality. Second, we verify that the function achieving this lower bound is indeed sufficient to satisfy the validity condition for any general distribution of $P^a$.

	We first show the necessity. Fix any $x \in [0, 1]$ such that $a(x) \le 1$. Consider a point-mass distribution for the auxiliary statistic, $P^a \equiv x$. In this case, Condition \eqref{eq:p-cond-1} must hold for all $s \in [0, 1]$. Specifically, choosing $s = a(x)$ (which is valid since $a(x) \le 1$), the condition becomes:
	\[
	\E \left[(1 - h(P^a))\mathbb{I}\{a(P^a) \le a(x)\}\right] = (1 - h(x)) \cdot 1 \le \beta a(x).
	\]
	This inequality implies $a(x) \ge (1 - h(x)) / \beta$.
	
	Next, we show the sufficiency of the choice $a(x) = (1 - h(x)) / \beta$. Substituting this form into the left-hand side of \eqref{eq:p-cond-1}, we have:
	\[
	\E \left[(1 - h(P^a))\mathbb{I}\left\{\frac{1 - h(P^a)}{\beta} \le s\right\}\right] = \E \left[(1 - h(P^a))\mathbb{I}\{1 - h(P^a) \le \beta s\}\right] \le \beta s.
	\]
\end{proof}

\subsection{Proof of Theorem \ref{thm:active_p_admissible}}

\begin{proof}
{
We consider non-negative functions $b$ and require \eqref{eq:p-cond-2} for every input law in the stated dependence class.

\par\smallskip\noindent\textit{Part 1: independence.}\enspace
Fix $q\in[0,1]$, take $P^a\equiv q$, and let $P\sim\mathrm{Uniform}(0,1)$. If $h(q)>0$, then $b(q)=0$ violates \eqref{eq:p-cond-2} at $s=0$. Thus $b(q)>0$, and choosing $0<s<\min\{1,b(q)\}$ gives
\[
h(q)\frac{s}{b(q)}\le(1-\beta)s,
\qquad\text{so}\qquad b(q)\ge\frac{h(q)}{1-\beta}.
\]
If $h(q)=0$, the integrand in \eqref{eq:p-cond-2} is zero, and the smallest non-negative choice is $b(q)=0$.

To prove sufficiency, take $b(q)=h(q)/(1-\beta)$. Conditional on $P^a=q$, the queried contribution is zero when $h(q)=0$. When $h(q)>0$, independence and super-uniformity give
\[
h(q)\pr\left(P\le\frac{(1-\beta)s}{h(q)}\right)
\le\min\{h(q),(1-\beta)s\}\le(1-\beta)s.
\]
Averaging over $P^a$ establishes \eqref{eq:p-cond-2}. Thus the proposed choice attains the necessary point-wise lower bound.

\par\smallskip\noindent\textit{Part 2: general dependence.}\enspace
Let $M:=\sup_x h(x)$ and $b^*(q):=M\mathbb{I}\{h(q)>0\}/(1-\beta)$. If $M=0$, the queried contribution is identically zero and $b^*\equiv0$ is point-wise smallest among non-negative functions. Hence assume $M>0$.

For validity, the factor $h(P^a)$ makes every contribution with $h(P^a)=0$ vanish. Therefore, for any joint law of $(P,P^a)$ with valid $P$,
\begin{align*}
\E[h(P^a)\mathbb{I}\{b^*(P^a)P\le s\}]
&=\E\left[h(P^a)\mathbb{I}\left\{\frac{M}{1-\beta}P\le s\right\}\right]\\
&\le M\pr\left(P\le\frac{(1-\beta)s}{M}\right)\\
&\le M\min\left\{1,\frac{(1-\beta)s}{M}\right\}
\le(1-\beta)s.
\end{align*}

For admissibility, suppose a non-negative function $\tilde b$ satisfies \eqref{eq:p-cond-2} for all such joint laws, with $\tilde b(q)\le b^*(q)$ everywhere and a strict inequality at $q_0$. Necessarily $h(q_0)>0$, since otherwise $b^*(q_0)=0$. Constant auxiliary statistics are allowed in this dependence class, so the necessary bound from Part 1 also applies to $\tilde b$. Consequently, for some $\delta\in(0,M)$,
\[
0<h(q_0)\le(1-\beta)\tilde b(q_0)=M-\delta<M.
\]
Choose
\[
0<\epsilon<\min\left\{\delta,\frac{h(q_0)\delta}{M-\delta}\right\},
\]
and, by the definition of $M$, choose $q_1$ with $h(q_1)>M-\epsilon$. Then $h(q_1)>M-\delta\ge h(q_0)$, so $q_1\ne q_0$.

Choose $0<s<\min\{1,(M-\delta)/(1-\beta)\}$ and set $p=(1-\beta)s/M\in(0,1)$. Take $P\sim\mathrm{Uniform}(0,1)$ and define $P^a=q_1$ when $P\le p$, and $P^a=q_0$ otherwise. On the first interval, $\tilde b(q_1)P\le Mp/(1-\beta)=s$. On the second interval,
\[
p<\frac{s}{\tilde b(q_0)}=\frac{(1-\beta)s}{M-\delta}<1.
\]
Thus
\begin{align*}
\E[h(P^a)\mathbb{I}\{\tilde b(P^a)P\le s\}]
&=p h(q_1)+h(q_0)\left\{\frac{(1-\beta)s}{M-\delta}-p\right\}\\
&>p(M-\epsilon)+p\frac{h(q_0)\delta}{M-\delta}\\
&=(1-\beta)s+p\left\{\frac{h(q_0)\delta}{M-\delta}-\epsilon\right\}\\
&>(1-\beta)s.
\end{align*}
This contradicts \eqref{eq:p-cond-2}, proving admissibility.
}
\end{proof}

\subsection{Proof of Proposition \ref{prop:no-optimal-h}}
\begin{proof}
	{Throughout the proof, $\beta$ is fixed and competing constructions use the same input statistics and the same $U\sim\mathrm{Uniform}(0,1)$, independent of those statistics. We first prove admissibility for $e$-values, then establish both directions of the characterisations for $p$-values.}
	
	\par\smallskip\noindent{\textit{Part 1: $e$-value setting.}}\enspace 
	Let $h_1$ and $h_2$ be two distinct control functions. Since they are distinct, there must exist a point $x_0 \ge 0$ where their values differ. Without loss of generality, assume $h_1(x_0) > h_2(x_0)$.
	
	To show that neither function can dominate the other, we {analyse} a simple setting where the auxiliary statistic is fixed: $\pr (E^a = x_0) = 1$. Let $E_1^{\text{active}}$ and $E_2^{\text{active}}$ be the active $e$-values generated using $h_1$ and $h_2$, respectively. The outcome depends on the draw of $U \sim \mathrm{Uniform}(0,1)$.
	
	{Consider the event $h_2(x_0)\le U<h_1(x_0)$, which has positive probability. Here $E_1^{\mathrm{active}}=(1-\beta)E/h_1(x_0)$ and $E_2^{\mathrm{active}}=\beta/(1-h_2(x_0))$. Taking $E\equiv0$ makes the second output strictly larger. Conversely, choose $L>\max\{1,\beta h_1(x_0)/[(1-\beta)(1-h_2(x_0))]\}$ and let $E=L$ with probability $1/L$ and $E=0$ otherwise. This is a valid $e$-value and makes the first output strictly larger with positive probability. Thus neither function dominates the other, including when $h_1(x_0)=1$ or $h_2(x_0)=0$, and every $h$ is admissible.}
	
{

\par\smallskip\noindent\textit{Part 2: $p$-value setting.}\enspace
All comparisons in this part are after truncation at one. We first show that any control function $\tilde h$ whose output dominates that of $h$ must satisfy $\tilde h(x)\ge h(x)$ for every $x$. If instead $\tilde h(x_0)<h(x_0)$, take $P^a\equiv x_0$ and $P\sim\mathrm{Uniform}(0,1)$. On the event $\tilde h(x_0)\le U<h(x_0)$, the construction using $h$ queries $P$, whereas the construction using $\tilde h$ returns the strictly positive value $\min\{1,(1-\tilde h(x_0))/\beta\}$. In either construction, the queried scaling coefficient for $h$ is finite and positive. Thus, for sufficiently small $P$, its output is strictly smaller with positive probability, contradicting domination. These inputs satisfy the independence assumption and are also allowed under general dependence.

\par\smallskip\noindent\textit{Independence.}\enspace
Suppose $h(x)>0$ for every $x$. By the preceding argument, a dominating $\tilde h$ must satisfy $\tilde h\ge h$. If $\tilde h\ne h$, choose $x_0$ such that $\tilde h(x_0)>h(x_0)>0$. With $P^a\equiv x_0$ and $P\sim\mathrm{Uniform}(0,1)$, the event $U<h(x_0)$ has positive probability and both constructions query $P$. For $0<P<\min\{1,(1-\beta)/\tilde h(x_0)\}$, their outputs satisfy
\[
\frac{h(x_0)}{1-\beta}P
<\frac{\tilde h(x_0)}{1-\beta}P<1.
\]
The strict comparison survives truncation and rules out domination. Hence $h$ is admissible.

Conversely, if $h(x_0)=0$, define $\tilde h$ to equal $h$ except that $\tilde h(x_0)=(1-\beta)/2$. At this auxiliary value, the original truncated output is always one, whereas the new truncated output is at most one. At every other auxiliary value the outputs coincide. Taking $P^a\equiv x_0$ and $P\sim\mathrm{Uniform}(0,1)$ gives a strict improvement with positive probability on the new query branch. Therefore $\tilde h$ dominates $h$, proving necessity.

\par\smallskip\noindent\textit{General dependence.}\enspace
Write $M=\sup_x h(x)$ and $\tilde M=\sup_x\tilde h(x)$. If $h\not\equiv0$ and $\tilde h$ dominates $h$, then $\tilde h\ge h$ implies $\tilde M\ge M$. In fact, equality is necessary. If $\tilde M>M$, choose $x_0$ with $h(x_0)>0$, fix $P^a\equiv x_0$ and take $P\sim\mathrm{Uniform}(0,1)$. On $U<h(x_0)$, both constructions query $P$, and sufficiently small $P$ gives
\[
\frac{M}{1-\beta}P
<\frac{\tilde M}{1-\beta}P<1,
\]
contradicting domination. In particular, if $h\equiv t$ for $t\in(0,1]$, the requirements $\tilde h\ge h$ and $\tilde M=M=t$ force $\tilde h=h$. Every positive constant function is therefore admissible.

Now suppose $h(x)>1-\beta$ for every $x$. Any distinct dominating function must satisfy $\tilde h\ge h$ and $\tilde h(x_0)>h(x_0)$ at some $x_0$. Take $P^a\equiv x_0$ and $P\equiv1$, a valid $p$-value. On $h(x_0)\le U<\tilde h(x_0)$, the construction using $h$ returns $(1-h(x_0))/\beta<1$, whereas the construction using $\tilde h$ queries $P$ and returns one after truncation, since $\tilde M\ge\tilde h(x_0)>1-\beta$. This contradicts domination and proves sufficiency of the strict lower bound.

It remains to prove necessity. The function $h\equiv0$ is dominated by the positive constant function $\tilde h\equiv(1-\beta)/2$, because its original truncated output is always one and the new query branch gives a strict improvement for a uniform $P$. If $h$ is nonconstant and does not satisfy the strict lower bound, there exists $x_0$ with $h(x_0)\le1-\beta$ and $h(x_0)<M$. Indeed, if a point satisfying the first inequality attains $M$, nonconstancy supplies a point with a smaller value. Define $\tilde h$ to equal $h$ except that $\tilde h(x_0)=M$. Then $\tilde M=M$, so the queried scaling coefficient is unchanged. At $x_0$, the original truncated output is one whenever $U\ge h(x_0)$, and increasing the query probability cannot make this output larger. On the common query branch and at every other auxiliary value, the outputs coincide. Finally, take $P^a\equiv x_0$ and $P\sim\mathrm{Uniform}(0,1)$. On $h(x_0)\le U<M$, sufficiently small $P$ makes the new queried output strictly smaller than one with positive probability. Thus $\tilde h$ dominates $h$. This argument does not require $M$ to be attained by $h$ and completes the characterisation.
}
\end{proof}


\subsection{Proof of Proposition \ref{prop:no-optimal-beta}}
\begin{proof}
	{Throughout, $h$ is fixed and competing choices of $\beta$ use the same input statistics and the same $U\sim\mathrm{Uniform}(0,1)$, independent of those statistics. For the $e$-value comparisons below, take $E\equiv1$ in each input law.}

	\par\smallskip\noindent{\textit{Part 1: $e$-value setting}}\enspace 
{Let $\beta_1>\beta_2$ be distinct values in $(0,1)$, and write $E_j^{\mathrm{active}}$ for the output using $\beta_j$. Since $h\not\equiv1$, choose $x$ with $h(x)<1$. With $E^a\equiv x$, on the event $U\ge h(x)$ we have $E_1^{\mathrm{active}}=\beta_1/(1-h(x))>\beta_2/(1-h(x))=E_2^{\mathrm{active}}$. Since $h\not\equiv0$, choose $y$ with $h(y)>0$. With $E^a\equiv y$, on the event $U<h(y)$ we have $E_1^{\mathrm{active}}=(1-\beta_1)/h(y)<(1-\beta_2)/h(y)=E_2^{\mathrm{active}}$. Both events have positive probability under their respective valid input laws, so neither choice dominates the other. The points $x$ and $y$ need not coincide.}

	\par\smallskip\noindent{\textit{Part 2: $p$-value setting.}}\enspace 
{
All comparisons are after truncation at one. Fix $\beta\in(0,1)$. First suppose that $1-\beta<h(x_0)<1$ for some $x_0$. Take $P^a\equiv x_0$ and $P\sim\mathrm{Uniform}(0,1)$, independently of $U$. These inputs are permitted under both independence and general dependence.

For any $\tilde\beta<\beta$, the event $U\ge h(x_0)$ has positive probability, and the non-query outputs satisfy
\[
\min\{1,P^{\mathrm{active}}_{h,\beta}\}
=\frac{1-h(x_0)}{\beta}
<\min\left\{1,\frac{1-h(x_0)}{\tilde\beta}\right\}
=\min\{1,P^{\mathrm{active}}_{h,\tilde\beta}\}.
\]
The strict inequality follows from $0<(1-h(x_0))/\beta<1$. Thus no smaller choice of $\beta$ can dominate the given one.

For any $\tilde\beta>\beta$, write the queried outputs as $CP/(1-\beta)$ and $CP/(1-\tilde\beta)$, where $C=h(x_0)>0$ under independence and $C=\sup_x h(x)>0$ under general dependence. The event
\[
U<h(x_0),\qquad 0<P<\min\{1,(1-\tilde\beta)/C\}
\]
has positive probability, and on this event
\[
\min\{1,P^{\mathrm{active}}_{h,\beta}\}
=\frac{CP}{1-\beta}
<\frac{CP}{1-\tilde\beta}
=\min\{1,P^{\mathrm{active}}_{h,\tilde\beta}\}<1.
\]
Thus no larger choice can dominate either, proving sufficiency.

Conversely, suppose that no $x$ satisfies $1-\beta<h(x)<1$, and choose any $\tilde\beta\in(0,\beta)$. Whenever $h(x)<1$, we have $1-h(x)\ge\beta>\tilde\beta$, so both non-query outputs equal one after truncation. Whenever a query occurs, its scaling numerator is unchanged by the choice of $\beta$, and $1-\tilde\beta>1-\beta$. The output using $\tilde\beta$ is therefore no larger for every valid input law. Since $h\not\equiv0$, choose $x_1$ with $h(x_1)>0$, take $P^a\equiv x_1$ and $P\sim\mathrm{Uniform}(0,1)$, independently of $U$. Write $C=h(x_1)>0$ under independence and $C=\sup_x h(x)>0$ under general dependence. On the positive-probability event $U<h(x_1)$ and $0<P<\min\{1,(1-\beta)/C\}$, the queried outputs satisfy $CP/(1-\tilde\beta)<CP/(1-\beta)<1$. Thus $\tilde\beta$ dominates $\beta$, proving necessity.
}
\end{proof}

\subsection{Proof of Proposition \ref{prop:systematic_sampling}}
\begin{proof}
The proof proceeds as follows: first, we show that $C_i \in \{0, 1\}$; second, we verify $\E [C_i] = p_i$; and finally, we demonstrate that the exact budget constraint $\sum_{i=1}^N C_i = n_b$ holds.
\par\smallskip\noindent{\textit{Support of $C_i$.}}\enspace 
By definition, $S_i - S_{i-1} = p_i \in [0, 1]$. Let $x = S_{i-1} - U$. We can rewrite the indicator as
\begin{equation*}
    C_i = \lfloor x + p_i \rfloor - \lfloor x \rfloor.
\end{equation*}
Since $0 \le p_i \le 1$, it follows that $x \le x + p_i \le x + 1$. The monotonicity of the floor function implies $\lfloor x \rfloor \le \lfloor x + p_i \rfloor \le \lfloor x \rfloor + 1$. Consequently, $0 \le C_i \le 1$. Because $C_i$ is defined as the difference between two integers, it must hold that $C_i \in \{0, 1\}$.

\par\smallskip\noindent{\textit{Expectation $\E [C_i] = p_i$.}}\enspace 
Consider the expectation of $\lfloor c - U \rfloor$ for an arbitrary constant $c \in \mathbb{R}$ and $U \sim \mathrm{Uniform}(0, 1)$. We decompose $c$ into its integer and fractional parts: $c = \lfloor c \rfloor + \{c\}$, where $\{c\} \in [0, 1)$. The random variable $\lfloor c - U \rfloor$ evaluates to $\lfloor c \rfloor$ if $U \le \{c\}$, and to $\lfloor c \rfloor - 1$ if $U > \{c\}$. Its expectation is therefore
\begin{align*}
    \E [\lfloor c - U \rfloor] &= \lfloor c \rfloor \cdot \pr (U \le \{c\}) + (\lfloor c \rfloor - 1) \cdot \pr (U > \{c\}) \\
    &= \lfloor c \rfloor \{c\} + (\lfloor c \rfloor - 1) (1 - \{c\}) \\
    &= \lfloor c \rfloor - 1 + \{c\} \\
    &= c - 1.
\end{align*}
So we have
\begin{equation*}
    \E [C_i] = \E [\lfloor S_i - U \rfloor] - \E [\lfloor S_{i-1} - U \rfloor] = (S_i - 1) - (S_{i-1} - 1) = S_i - S_{i-1} = p_i.
\end{equation*}

\par\smallskip\noindent{\textit{Exact sum constraint.}}\enspace 
Summing $C_i$ over all $N$ variables yields:
\begin{equation*}
    \sum_{i=1}^N C_i = \sum_{i=1}^N \left( \lfloor S_i - U \rfloor - \lfloor S_{i-1} - U \rfloor \right) = \lfloor S_N - U \rfloor - \lfloor S_0 - U \rfloor.
\end{equation*}
By construction, $S_0 = 0$ and $S_N = \sum_{j=1}^N p_j = n_b \in \mathbb{N}$. Then we have
\begin{equation*}
    \sum_{i=1}^N C_i = \lfloor n_b - U \rfloor - \lfloor -U \rfloor = n_b + \lfloor -U \rfloor - \lfloor -U \rfloor = n_b.
\end{equation*}
\end{proof}

{
\begingroup
\raggedbottom
\subsection{Proof of Theorem~\ref{thm:active_bh_conditional}}
Let $R$ be the number of BH rejections and write $\mathbf C=(C_1,\ldots,C_N)$. We decompose the FDR according to whether each true null is queried:
\[
{\text{\textsc{fdr}}}
=\sum_{i\in\mathcal H_0}\left\{
\E \left[\frac{(1-C_i)\mathbb I\{i\text{ is rejected}\}}{R\vee1}\right]
+\E \left[\frac{C_i\mathbb I\{i\text{ is rejected}\}}{R\vee1}\right]
\right\}.
\]
For each $i\in\mathcal H_0$, we bound the nonquery and query contributions by $\alpha\beta/N$ and $\alpha(1-\beta)/N$, respectively. Their sum is at most $\alpha/N$.

For each $i\in\mathcal H_0$, use the queried coefficient $b_i$ in the theorem. Both choices satisfy $h_i/b_i\le1-\beta$ on $\{h_i>0\}$. A term corresponding to $C_i=1$ when $h_i=0$, or to $C_i=0$ when $h_i=1$, is zero almost surely and is omitted below. Since every BH threshold is at most $\alpha<1$, clipping the active values at one does not change any rejection event.

For the nonquery branch, the factor $1-C_i$ makes the contribution zero when $C_i=1$. When $C_i=0$, the active $p$-value is $\min\{1,(1-h_i)/\beta\}$, so BH rejects $i$ if and only if $R>0$ and $\min\{1,(1-h_i)/\beta\}\le\alpha R/N$. Since $\alpha R/N<1$, this is equivalent to $R>0$ and $(1-h_i)/\beta\le\alpha R/N$, giving the first equality below. On this rejection event, $R\vee1=R$ and $1/R\le\alpha\beta/\{N(1-h_i)\}$ whenever $h_i<1$. When $h_i=1$, $C_i=1$ almost surely, so the contribution is zero. Hence
\[
\begin{aligned}
\E \left[
\frac{(1-C_i)\mathbb I\{i\text{ is rejected}\}}{R\vee1}
\right]
&=\E \left[
\frac{(1-C_i)\mathbb I\{(1-h_i)/\beta\le\alpha R/N,\ R>0\}}{R\vee1}
\right]\\
&\le\frac{\alpha\beta}{N}\E \left[
\frac{1-C_i}{1-h_i}\mathbb I\{h_i<1\}\right]\\
&=\frac{\alpha\beta}{N}\E \left[
\E \left[\left.
\frac{1-C_i}{1-h_i}\mathbb I\{h_i<1\}
\right|\mathbf P^a\right]\right]\\
&=\frac{\alpha\beta}{N}\E \left[
\frac{1-h_i}{1-h_i}\mathbb I\{h_i<1\}\right]\\
&=\frac{\alpha\beta}{N}\pr (h_i<1)
\le\frac{\alpha\beta}{N}.
\end{aligned}
\]

For the query branch, our goal is to show that
\[
\E \left[\frac{C_i\mathbb I\{i\text{ is rejected}\}}{R\vee1}\right]
\le\frac{\alpha(1-\beta)}{N}.
\]
Here the rejection threshold depends on $R$, which itself depends on $P_i$. The following lemma controls this dependence.
\begin{lemma}
\label{lem:random_threshold_prds}
Let $W\in[0,1]$ be superuniform and let $K\in\{0,\ldots,N\}$. Suppose that, for each $r=1,\ldots,N$, the function $t\mapsto\pr (K\ge r\mid W\le t)$ is nonincreasing wherever the conditioning event has positive probability. Then, for every $c>0$,
\[
\E \left[\frac{\mathbb I\{W\le cK,\ K>0\}}{K\vee1}\right]\le c.
\]
\end{lemma}

\begin{proof}
Write $F(t)=\pr (W\le t)$, $g_r(t)=\pr (K\ge r\mid W\le t)$ and $t_r=\min(cr,1)$. Set $g_{N+1}=0$ and $t_{N+1}=1$. If $F(t_N)=0$, the claimed expectation is zero. Otherwise, let $r_0$ be the first index with $F(t_{r_0})>0$.

Conditioning on each possible value $K=r$ gives the first equality below. Superuniformity gives $F(t_r)/r\le c$, while the assumed monotonicity gives $g_{r+1}(t_r)\ge g_{r+1}(t_{r+1})$. These yield the two inequalities, respectively:
\begin{align*}
\E \left[\frac{\mathbb I\{W\le cK,\ K>0\}}{K\vee1}\right]
&=\sum_{r=r_0}^N\frac{F(t_r)}{r}
\{g_r(t_r)-g_{r+1}(t_r)\}\\
&\le c\sum_{r=r_0}^N\{g_r(t_r)-g_{r+1}(t_r)\}\\
&\le c\sum_{r=r_0}^N\{g_r(t_r)-g_{r+1}(t_{r+1})\}\\
&=c\,g_{r_0}(t_{r_0})\le c.
\end{align*}
The last sum telescopes. All conditional probabilities used above are defined on events of positive probability.
\end{proof}

We now apply Lemma~\ref{lem:random_threshold_prds} conditional on $(\mathbf P^a,\mathbf C)$. Since $\mathbf C\perp\mathbf P\mid\mathbf P^a$, conditioning additionally on $\mathbf C$ preserves the assumed superuniformity and PRDS of $\mathbf P$ conditional on $\mathbf P^a$.

With $(\mathbf P^a,\mathbf C)$ fixed, increasing any original $p$-value cannot decrease an active $p$-value and hence cannot increase $R$. Therefore $\{R<r\}$ is an increasing set in $\mathbf P$. Conditional PRDS makes its probability given $P_i\le t$ nondecreasing in $t$. Taking complements shows that $t\mapsto\pr (R\ge r\mid P_i\le t,\mathbf P^a,\mathbf C)$ is nonincreasing, as required by the lemma.

On $\{C_i=1\}$, BH rejects $i$ exactly when $P_i\le\alpha R/(Nb_i)$ and $R>0$. Applying the lemma with $W=P_i$, $K=R$ and $c=\alpha/(Nb_i)$ gives
\[
\E \left[\left.
\frac{C_i\mathbb I\{i\text{ is rejected}\}}{R\vee1}
\right|\mathbf P^a,\mathbf C\right]
\le\frac{\alpha C_i}{Nb_i}.
\]
On $\{C_i=0\}$, both sides are zero, with the ratio defined as zero if $b_i=0$. Finally, the law of iterated expectations and $\E [C_i\mid\mathbf P^a]=h_i$ give
\[
\begin{aligned}
\E \left[\frac{C_i\mathbb I\{i\text{ is rejected}\}}{R\vee1}\right]
&\le\frac{\alpha}{N}\E \left[
\frac{C_i}{b_i}\mathbb I\{h_i>0\}\right]\\
&=\frac{\alpha}{N}\E \left[
\E \left[\left.\frac{C_i}{b_i}\mathbb I\{h_i>0\}\right|\mathbf P^a\right]\right]\\
&=\frac{\alpha}{N}\E \left[
\frac{h_i}{b_i}\mathbb I\{h_i>0\}\right]
\le\frac{\alpha(1-\beta)}{N},
\end{aligned}
\]
where the last inequality uses $h_i/b_i\le1-\beta$ on $\{h_i>0\}$.
The two branches therefore contribute at most $\alpha/N$ for each true null. Summing over $i\in\mathcal H_0$ proves ${\text{\textsc{fdr}}}\le\alpha|\mathcal H_0|/N$.

\par
\endgroup

\subsection{Proof of Corollary~\ref{cor:active_bh_independent}}
\begin{proof}
Independence of $\mathbf P$ and $\mathbf P^a$ implies that the conditional law of the full vector $\mathbf P$ given $\mathbf P^a$ equals its marginal law almost surely. Hence marginal validity of each null $P_i$ gives conditional validity, and marginal PRDS of $\mathbf P$ gives conditional PRDS. The construction and sampling assumptions are unchanged, so Theorem~\ref{thm:active_bh_conditional} applies.
\end{proof}
}

\subsection{Proof of Proposition \ref{prop:multi}}
\begin{proof}
{
Fix $\boldsymbol\beta$ and $\mathbf h\in\mathcal H$. Any distinct competing allocation $\tilde{\mathbf h}\in\mathcal H$ must satisfy $h_i(\mathbf x)>\tilde h_i(\mathbf x)$ for some auxiliary vector $\mathbf x$ and coordinate $i$, since both allocations sum to $n_b$ at every input. Fix $\mathbf X^a\equiv\mathbf x$ and generate the query indicators $\mathbf C$ and $\tilde{\mathbf C}$ using the same $U$ as in Algorithm~\ref{alg:unified_framework}. Their marginal probabilities imply
\[
\pr(C_i=1,\tilde C_i=0)
=h_i(\mathbf x)-\pr(C_i=1,\tilde C_i=1)
\ge h_i(\mathbf x)-\tilde h_i(\mathbf x)>0.
\]
Thus the original allocation queries coordinate $i$ while the competing allocation does not, with positive probability. The argument uses only the query marginals and does not require independent or nested query events.

\par\smallskip\noindent\textit{Part 1: $e$-values.}\enspace
Choose
\[
L>\max\left\{1,\frac{\beta_i h_i(\mathbf x)}{(1-\beta_i)(1-\tilde h_i(\mathbf x))}\right\},
\]
and let $E_i=L$ with probability $1/L$ and $E_i=0$ otherwise, independently of $U$. Set all other original $e$-values equal to one. These inputs are valid since $\E[E_j]=1$ for every $j$. On the event $C_i=1$, $\tilde C_i=0$ and $E_i=L$, which has positive probability, the original and competing outputs satisfy
\[
\frac{(1-\beta_i)L}{h_i(\mathbf x)}
>\frac{\beta_i}{1-\tilde h_i(\mathbf x)}.
\]
The denominators are positive because $0\le\tilde h_i(\mathbf x)<h_i(\mathbf x)\le1$. Hence the competing vector cannot dominate the original vector.

\par\smallskip\noindent\textit{Part 2: $p$-values.}\enspace
Take $P_i\sim\mathrm{Uniform}(0,1)$ independently of $U$, and set all other original $p$-values equal to one. The deterministic auxiliary vector is independent of the full original vector, so these inputs are allowed under both constructions. On $C_i=1$ and $\tilde C_i=0$, the competing output after truncation is $r=\min\{1,(1-\tilde h_i(\mathbf x))/\beta_i\}>0$. The original queried scale is $b=h_i(\mathbf x)/(1-\beta_i)>0$ under independence and $b=\sup_{\mathbf y}h_i(\mathbf y)/(1-\beta_i)>0$ under general dependence. Both scales are finite. On the positive-probability event
\[
C_i=1,\qquad \tilde C_i=0,\qquad 0<P_i<\min\{1,r/b\},
\]
the original output satisfies $\min\{1,bP_i\}<r$. Thus the competing vector cannot dominate the original vector after truncation at one.

Since every distinct competing allocation is excluded in both settings, every $\mathbf h\in\mathcal H$ is admissible with $\boldsymbol\beta$ fixed. When $n_b=0$ or $n_b=N$, the class $\mathcal H$ contains only one allocation, and the conclusion is immediate.
}
\end{proof}

\subsection{Proof of Proposition~\ref{prop:multi-beta-adm}}
\begin{proof}
{
Fix $\mathbf h$ satisfying the stated assumptions and $\boldsymbol\beta\in(0,1)^N$. Since $\mathbf h$ is fixed, Algorithm~\ref{alg:unified_framework} uses the same query indicators for all competing hyperparameter vectors: $\mathbf C$ is determined by $\mathbf X^a$ and $U$ and does not depend on $\boldsymbol\beta$. Conditional on $\mathbf X^a=\mathbf x$, $C_i$ has probability $h_i(\mathbf x)$ of being one and is independent of the original statistics. All witness input laws below take the auxiliary vector to be deterministic and the original statistics to be independent of $U$.

\par\smallskip\noindent\textit{Part 1: $e$-values.}\enspace
Take any competing vector $\tilde{\boldsymbol\beta}\ne\boldsymbol\beta$ and choose a coordinate $i$ where they differ. Set every original $e$-value equal to one. If $\tilde\beta_i<\beta_i$, choose $\mathbf x$ with $h_i(\mathbf x)<1$, which exists because $h_i\not\equiv1$. With $\mathbf E^a\equiv\mathbf x$, the event $C_i=0$ has positive probability and the original output is strictly larger:
\[
\frac{\beta_i}{1-h_i(\mathbf x)}
>\frac{\tilde\beta_i}{1-h_i(\mathbf x)}.
\]
If $\tilde\beta_i>\beta_i$, choose $\mathbf x$ with $h_i(\mathbf x)>0$, which exists because $h_i\not\equiv0$. The event $C_i=1$ then has positive probability and
\[
\frac{1-\beta_i}{h_i(\mathbf x)}
>\frac{1-\tilde\beta_i}{h_i(\mathbf x)}.
\]
In either case, the competing vector fails the required component-wise inequality. Hence every $\boldsymbol\beta$ is admissible for active $e$-values.

\par\smallskip\noindent\textit{Part 2: $p$-values.}\enspace
All comparisons are after truncation at one. First suppose that, for each $i$, there exists $\mathbf x$ such that $1-\beta_i<h_i(\mathbf x)<1$. Take any competing vector $\tilde{\boldsymbol\beta}\ne\boldsymbol\beta$, select a coordinate $i$ where they differ, and fix $\mathbf P^a\equiv\mathbf x$ at a witness for this coordinate. Different coordinates need not have the same witness. Let $P_i\sim\mathrm{Uniform}(0,1)$ and set $P_j\equiv1$ for $j\ne i$. These valid inputs satisfy the independence requirements of both constructions.

If $\tilde\beta_i<\beta_i$, then on $C_i=0$, an event of positive probability, the original output is strictly smaller than the competing output:
\[
\frac{1-h_i(\mathbf x)}{\beta_i}
<\min\left\{1,\frac{1-h_i(\mathbf x)}{\tilde\beta_i}\right\},
\]
because $0<(1-h_i(\mathbf x))/\beta_i<1$. If $\tilde\beta_i>\beta_i$, write the queried numerators as $M=h_i(\mathbf x)>0$ under independence and $M=\sup_{\mathbf y}h_i(\mathbf y)>0$ under general dependence. This numerator is the same for both choices of hyperparameters. On the positive-probability event $C_i=1$ and $0<P_i<\min\{1,(1-\tilde\beta_i)/M\}$,
\[
\frac{MP_i}{1-\beta_i}
<\frac{MP_i}{1-\tilde\beta_i}<1.
\]
Thus the competing vector cannot dominate in either case, proving sufficiency.

For necessity, suppose that for some coordinate $i$ no auxiliary vector satisfies $1-\beta_i<h_i(\mathbf x)<1$. Define $\tilde\beta_i=\beta_i/2$ and $\tilde\beta_j=\beta_j$ for $j\ne i$. Every possible non-query branch in coordinate $i$ has $h_i(\mathbf x)<1$ and hence $1-h_i(\mathbf x)\ge\beta_i$, so its output is one after truncation under both hyperparameter vectors. On the query branch, decreasing $\beta_i$ decreases the scale, and all other coordinates are unchanged. Consequently, the competing vector is component-wise no larger for every valid input law.

To obtain strict improvement, choose $\mathbf x$ with $h_i(\mathbf x)>0$, which exists by non-degeneracy, and use the same type of input law as above: $\mathbf P^a\equiv\mathbf x$, $P_i\sim\mathrm{Uniform}(0,1)$ and $P_j\equiv1$ for $j\ne i$. With $M$ defined for the relevant construction as above, the event $C_i=1$ and $0<P_i<\min\{1,(1-\beta_i)/M\}$ has positive probability, and
\[
\frac{MP_i}{1-\tilde\beta_i}
<\frac{MP_i}{1-\beta_i}<1.
\]
Thus $\tilde{\boldsymbol\beta}$ dominates $\boldsymbol\beta$, proving necessity. The final assertion follows because the supremum condition supplies the required auxiliary vector for every $i$ and every $\beta_i\in(0,1)$.
}
\end{proof}

\subsection{Proof of Lemma \ref{lem:calibrator_bound}}
\begin{proof}
	{Since $f$ transforms a uniform $p$-value into a valid $e$-value, $\int_0^1 f(x)\,dx\le1$. If there exists $x_0\in(0,1]$ such that $f(x_0)x_0>1$, then:}
	$$
	\int_{0}^{x_0} f(x)  dx \geq \int_{0}^{x_0} f(x_0)  dx = f(x_0) \cdot x_0 > 1
	$$
	This implies:
	$$
	\int_{0}^{1} f(x)  dx \geq \int_{0}^{x_0} f(x)  dx > 1
	$$
	which is a contradiction.
\end{proof}

\section{Counterexample for the Choice of $b(\cdot)$ in Theorem \ref{thm:active_p_admissible}}
\label{sec:counterexample}

{
The choice $b(q)=h(q)/(1-\beta)$ from the independent construction can produce an invalid active $p$-value when $P$ and $P^a$ are dependent. We show this by computing the tail probability of the complete active statistic.

\par\smallskip\noindent\textit{Construction.}\enspace
Take $\beta=1/10$ and $h(q)=(3+7q)/10$ for $q\in[0,1]$. Let $P^a$ take the values zero and one with equal probability, and let
\[
P\mid(P^a=0)\sim\mathrm{Uniform}(1/2,1),
\qquad
P\mid(P^a=1)\sim\mathrm{Uniform}(0,1/2).
\]
The marginal distribution of $P$ is $\mathrm{Uniform}(0,1)$, so $P$ is a valid null $p$-value. Let $U\sim\mathrm{Uniform}(0,1)$ be independent of $(P,P^a)$. Using $a(q)=(1-h(q))/\beta$ and $b(q)=h(q)/(1-\beta)$ in Definition~\ref{def:active-p-value} gives
\[
P^{\mathrm{active}}=
\begin{cases}
7 & \text{if }P^a=0,\ U\ge3/10,\\[0.4em]
P/3 & \text{if }P^a=0,\ U<3/10,\\[0.4em]
10P/9 & \text{if }P^a=1.
\end{cases}
\]

\par\smallskip\noindent\textit{Failure of Super-uniformity.}\enspace
Set $s=3/5$. When $P^a=0$, the query output satisfies $P/3\le1/3<s$, whereas the nonquery output is $7>s$. When $P^a=1$, a query always occurs and $10P/9\le5/9<s$. Consequently,
\[
\pr\bigl(P^{\mathrm{active}}\le3/5\bigr)
=\frac12\cdot\frac3{10}+\frac12\cdot1
=\frac{13}{20}>\frac35.
\]
The same inequality holds after truncation at one, since $s<1$. Thus the resulting statistic is not a valid $p$-value, showing that the independent construction cannot be used under arbitrary dependence without additional assumptions.
}


\section{Counterexample for the Decomposition in Remark \ref{rem:decomposition}}
\label{sec:counterexample-decomposition}

In our main analysis, we adopted a decomposition strategy, ensuring the validity of the active $p$-value by separately controlling the two components of its tail probability, as shown in equations \eqref{eq:p-cond-1} and \eqref{eq:p-cond-2}. A natural question arises: is this decomposition necessary? That is, for any valid active $p$-value construction satisfying the super-uniformity condition, must there exist a universal $\beta \in [0,1]$ that validates the decomposition?

We show that the answer is no. We construct a simple, valid active $p$-value for which no such universal $\beta$ can be found.

\par\smallskip\noindent{\textit{A Valid Construction That Defies Decomposition.}}\enspace 
Consider the specific construction where $a(p) \equiv 1$ and $b(p) \equiv 1$ for all $p \in [0,1]$. Let $h: [0,1] \to [0,1]$ be any non-constant function (e.g., $h(x)=x$). The active $p$-value is then:
\[
P^{\mathrm{active}} =
\begin{cases}
	1 & \text{if } U \geq h(P^a) \\
	P & \text{if } U < h(P^a)
\end{cases}
\]
where $U \sim \mathrm{Uniform}(0,1)$ is independent of $(P, P^a)$. This construction is demonstrably valid. For any $s \in [0,1)$, the proxy branch can never produce a value $\leq s$. Therefore,
\[
\pr (P^{\mathrm{active}} \leq s) = \pr (U < h(P^a) \text{ and } P \leq s) \leq \pr (P \leq s) \leq s.
\]
The super-uniformity condition holds for any valid $(P, P^a)$ distribution.

\par\smallskip\noindent{\textit{Deriving the Contradiction.}}\enspace 
Now, assume for the sake of contradiction that there exists a universal $\beta \in [0,1]$ for which the decomposition conditions \eqref{eq:p-cond-1} and \eqref{eq:p-cond-2} hold for this construction.

First, we {analyse} condition \eqref{eq:p-cond-1}. Since $a(P^a) \equiv 1$, the indicator $\mathbb{I}\{a(P^a) \le s\}$ is 0 for $s<1$ and 1 for $s=1$. The condition is trivially satisfied for $s<1$. For $s=1$, it requires:
\[
\E [(1 - h(P^a)) \cdot \mathbb{I}\{1 \le 1\}] = \E [1 - h(P^a)] \leq \beta \cdot 1.
\]
This inequality must hold for any distribution of $P^a$. If we choose a deterministic $P^a \equiv x$ for any $x \in [0,1]$, this implies $1 - h(x) \leq \beta$, which rearranges to a lower bound on $h(x)$:
\begin{equation}\label{eq:h_lower_bound}
	h(x) \geq 1 - \beta \quad \text{for all } x \in [0,1].
\end{equation}

Next, we {analyse} condition \eqref{eq:p-cond-2}. With $b(P^a) \equiv 1$, it states:
\[
\E [h(P^a) \cdot \mathbb{I}\{P \le s\}] \leq (1-\beta)s.
\]
To isolate $h(x)$, we can again choose a deterministic $P^a \equiv x$ and an independent $P \sim \mathrm{Uniform}(0,1)$. The condition becomes:
\[
h(x) \cdot \pr (P \le s) = h(x) \cdot s \leq (1-\beta)s.
\]
This must hold for all $s \in (0,1]$, which implies an upper bound on $h(x)$:
\begin{equation}\label{eq:h_upper_bound}
	h(x) \leq 1 - \beta \quad \text{for all } x \in [0,1].
\end{equation}

Combining the lower bound from \eqref{eq:h_lower_bound} and the upper bound from \eqref{eq:h_upper_bound}, we find that for a universal $\beta$ to exist, the function $h(x)$ must satisfy $h(x) = 1 - \beta$ for all $x \in [0,1]$. This means $h(x)$ must be a constant function.

This contradicts our initial premise that $h(x)$ is a non-constant function. Therefore, our assumption that a universal $\beta$ exists must be false. This counterexample confirms that the decomposition is a sufficient, but not necessary, condition for the validity of an active $p$-value.

\section{Comparison with the Framework of \texorpdfstring{\cite{xu2025active}}{Xu et al. (2025)}}
\label{sec:comparison_xu}

In this section, we formalize the relationship between our active testing framework and the closely related method of \citet{xu2025active}. We show that for $e$-values, our construction point-wise dominates their basic construction, {yielding at least as much power at any common rejection threshold for an identical computational cost}. For $p$-values, our construction is {no less powerful at any common rejection threshold} under independence, while under general dependence, the two frameworks are equivalent, revealing the \citet{xu2025active} construction to be a special case of ours.

\subsection{Comparison of $e$-value Constructions}

We begin by comparing the active $e$-value constructions. The \citet{xu2025active} method defines a query probability based on an auxiliary $e$-value $E^a$ and a hyperparameter $\beta \in (0,1)$. A Bernoulli random variable $T \sim \mathrm{Bern}((1 - \beta(E^a)^{-1})_{+})$ determines whether to query the exact $e$-value $E$. The final statistic is reported as:
\[
\tilde{E} := (1 - T)E^a + T(1 - \beta)E. \quad \text{(\citealp{xu2025active} construction)}
\]
To establish a direct comparison, we adopt the identical decision rule in our framework by setting the control function to $h(x) = (1 - \beta x^{-1})_{+}$. Our active $e$-value is then constructed as:
\[
E^{\mathrm{active}} :=
\begin{cases}
	\max\{\beta, E^a\} & \text{if } T=0 \\
	(1-\beta)\dfrac{E^a}{E^a-\beta}E & \text{if } T=1
\end{cases}. \quad \text{(Our construction)}
\]

\par\smallskip\noindent{\textit{Derivation of the Proxy-Branch Term.}}\enspace 
The $\max\{\beta, E^a\}$ term in our construction for the $T=0$ (proxy) branch arises directly from the optimal form of an active $e$-value given in Corollary \ref{cor:optimal-construction}, which is $\beta/(1-h(E^a))$. With our specific choice of $h(E^a) = (1-\beta(E^a)^{-1})_{+}$, we {analyse} the denominator in two cases:
\begin{enumerate}[label={(\roman*)}]
	\item If $E^a \le \beta$, then $1-\beta(E^a)^{-1} \le 0$, so $h(E^a)=0$. The term becomes $\beta/(1-0) = \beta$.
	\item If $E^a > \beta$, then $h(E^a) = 1-\beta(E^a)^{-1}$. The term becomes $\beta/\left(1 - (1-\beta(E^a)^{-1})\right) = \beta/\left(\beta(E^a)^{-1}\right) = E^a$.
\end{enumerate}
Combining these two cases, where the result is $\beta$ if $E^a \le \beta$ and $E^a$ if $E^a > \beta$, gives precisely $\max\{\beta, E^a\}$. 

\par\smallskip\noindent{\textit{Case 1: $E^a \le \beta$.}}\enspace  The query probability is {$h(E^a)=0$}, so $T=0$ almost surely. The resulting statistics are deterministic:
\[
\tilde{E} = E^a \quad \text{and} \quad E^{\mathrm{active}} = \beta.
\]
{Thus, $E^{\mathrm{active}} \ge \tilde{E}$, with strict inequality if $E^a < \beta$ and equality if $E^a=\beta$.}

\par\smallskip\noindent{\textit{Case 2: $E^a > \beta$.}}\enspace  Both outcomes for $T$ occur with positive probability. Conditional on $T$, the statistics are:
\[
\tilde{E} =
\begin{cases}
	E^a & \text{if } T=0 \\
	(1-\beta)E & \text{if } T=1
\end{cases}
\quad \text{and} \quad
E^{\mathrm{active}} =
\begin{cases}
	E^a & \text{if } T=0 \\
	(1-\beta)\dfrac{E^a}{E^a-\beta}E & \text{if } T=1
\end{cases}.
\]
{When $T=1$, the scaling factor satisfies $(E^a)/(E^a-\beta)>1$. Hence $E^{\mathrm{active}} \geq \tilde{E}$, with strict inequality if and only if $E>0$. When $T=0$, the two statistics are equal.}

In summary, our construction dominates the basic construction of \citet{xu2025active} point-wise:
\begin{enumerate}[label={(\roman*)}]
	\item {Almost sure inequality:} $E^{\mathrm{active}} \ge \tilde{E}$.
	\item {Strict improvement:} The inequality is strict whenever $E^a < \beta$. When $E^a > \beta$, it is strict on the event {$\{T=1,E>0\}$}.
\end{enumerate}

The comparison above concerns the basic active $e$-value construction in the first arXiv version of \citet{xu2025active}. Appendix E of their second arXiv version (submitted after the first arXiv version of our paper) introduces a boosted construction. With the same proxy, tuning parameter $\beta\in(0,1)$ and query indicator, their boosted construction is still pointwise no larger than construction~\eqref{active-e} with a specific control function $h(x)=(1-\beta/x)_+$, with equality when $E^a\geq\beta$. Their construction is a special case of our general framework. Beyond this construction, we establish optimality and admissibility results under the stated conditions and develop global allocation under an exact computational budget. These results are not established in the second arXiv version of their paper.

\subsection{Comparison of $p$-value Constructions}

Next, we compare the active $p$-value constructions. The \citet{xu2025active} method uses a query probability based on an auxiliary $p$-value $P^a$ and defines $T \sim \mathrm{Bern}((1 - \beta P^a)_{+})$. The final statistic is:
\[
\tilde{P} := (1 - T)P^a + T(1 - \beta)^{-1} P. \quad \text{(\citealp{xu2025active} construction)}
\]
We adopt the same decision rule by setting our control function $h(x) = (1 - \beta x)_{+}$ and letting $T := \mathbb{I}\{U < h(P^a)\}$ for $U \sim \mathrm{Uniform}(0,1)$.

\par\smallskip\noindent{\textit{Independent Setting.}}\enspace 
Our active $p$-value under independence is given by:
\[
P^{\mathrm{active}} = (1 - T)\frac{1 - h(P^a)}{\beta} + T\frac{h(P^a)}{1 - \beta}P.
\]
We compare this to $\tilde{P}$ in each branch of the random trial $T$.
\begin{enumerate}[label={(\roman*)}]
	\item Conditional on $T=0$: $\tilde{P} = P^a$. Our construction yields $P^{\mathrm{active}} = {(1 - h(P^a))/(\beta)} = {(1 - \max\{0, 1-\beta P^a\})/(\beta)} = \min\{\beta^{-1}, P^a\}$. Since $\beta \in (0,1)$, $\beta^{-1} > 1$, and since $P^a \in [0,1]$, it follows that $\min\{\beta^{-1}, P^a\} = P^a$. Thus, $P^{\mathrm{active}} = \tilde{P}$.
	\item Conditional on $T=1$: $\tilde{P} = {(P)/(1-\beta)}$. Our construction yields $P^{\mathrm{active}} = {(h(P^a))/(1 - \beta)}P = {(\max\{0, 1-\beta P^a\})/(1 - \beta)}P$. Since $\max\{0, 1-\beta P^a\} \leq 1$, we have $P^{\mathrm{active}} \leq \tilde{P}$, with strict inequality whenever $P^a>0$ and $P>0$.
\end{enumerate}
{Thus, $P^{\mathrm{active}}\le\tilde{P}$ point-wise, so our construction is no less powerful at any common rejection threshold under independence.}

\par\smallskip\noindent{\textit{General Dependence Setting.}}\enspace 
Our active $p$-value under general dependence takes the form:
\[
P^{\mathrm{active}} = (1 - T)\frac{1 - h(P^a)}{\beta} + T\frac{1}{1 - \beta}P.
\]
As shown above, the term for the $T=0$ branch simplifies to $P^a$. The term for the $T=1$ branch is identical to that of $\tilde{P}$. The entire expression is therefore:
\[
P^{\mathrm{active}} = (1-T)P^a + T(1-\beta)^{-1}P = \tilde{P}.
\]
The two constructions are identical. This reveals that the \citet{xu2025active} procedure arises as a special case of our more general framework when the conservative construction for arbitrary dependence is employed.


\end{document}